\documentclass[twoside]{elsart}
\usepackage{amsmath,latexsym,amssymb}
\journal{Annals of Physics}

\numberwithin{equation}{section}
\newcommand{\R}{\mathbb R}
\newcommand{\C}{\mathbb C}

\newcommand{\K}{\Bold{K}}
\newcommand{\E}{\mathbb E}

\newcommand{\p}{\partial}

\newcommand{\Xa}{\mathbf{X}_{(\alpha)}}
\newcommand{\Xb}{\mathbf{X}_{(\beta)}}

\newcommand{\Za}{{Z^d_a}}
\newcommand{\Z}{Z^d_{b}}

\newcommand{\Ta}{T^+_a}
\newcommand{\Tb}{T^+_b}

\newcommand{\PM}{P_a^{\mathbb{V}}\mathbb{M}}
\newcommand{\PMp}{P_{b}^{\mathbb{V}}\mathbb{M}}
\newcommand{\PU}{P_a^{\mathbb{V}}\mathbb{U}}

\newcommand{\DQ}[1]{\mathcal{D}_{{Q},{W}}#1}
\newcommand{\DO}[1]{\mathcal{D}#1}

\newcommand{\Bold}[1]{\mbox{\boldmath$\mathit{#1}$}}
\newcommand{\dual}{\langle \tau',\tau\rangle}

\newcommand{\taurange}{\int_{C_+}}
\newcommand{\ti}{\mathrm{t}}
\newcommand{\x}{\mathrm{x}}
\newcommand{\y}{\mathrm{y}}

\newcommand{\M}{\mathbb{M}}
\newcommand{\U}{\mathbb{U}}

\newcommand{\QED}{\hspace{.2in}\square\newline}

\newcommand{\taubot}{\tau_{\mathrm{x}_a}^\bot}
\newcommand{\taubotp}{\tau_{\mathrm{x}_{b}}^\bot}

\newcommand{\tr}{\tau}

\newtheorem{theorem}{Theorem}[section]
\newtheorem{corollary}{Corollary}[section]

\newtheorem{lemma}{Lemma}[section]
\begin{document}

\begin{frontmatter}
\title{Path Integral Solution of Linear Second Order Partial
Differential Equations II. Elliptic, Parabolic and Hyperbolic
Cases}
\author{J. LaChapelle}
\ead{jlachapelle@comcast.net}

\begin{abstract}
A theorem that constructs a path integral solution for general
second order partial differential equations is specialized to
obtain path integrals that are solutions of elliptic, parabolic,
and hyperbolic linear second order partial differential equations
with Dirichlet/Neumann boundary conditions. The construction is
checked by evaluating several known kernels for regions with
planar and spherical boundaries. Some new calculational techniques
are introduced.
\end{abstract}

\begin{keyword}
partial differential equations\sep path integrals\sep functional
integration \PACS 2.30.Cj\sep 2.30.Jr
\end{keyword}
\end{frontmatter}


\section{Introduction}
The general path integral developed in \cite{LA2} is specialized
to yield path integral solutions to elliptic, parabolic, and
hyperbolic linear second order PDEs. The examples in Section
\ref{sec. examples} provide explicit realizations of the
construction and can be compared to known solutions. Some new
calculational techniques are introduced which may offer some
advantage in numerical methods.

Material and notation presented in \cite{LA2} will be assumed
here. The reader requiring motivation to fully digest \cite{LA2}
may wish to begin with perhaps more familiar material in the
examples presented here in Subsections \ref{sec. boundary at
infinity} (the Poisson, diffusion/Schr\"{o}dinger, and wave
equations in unbounded space), \ref{sec. the half-space} (the
diffusion equation in the half-plane), and \ref{sec. the n-ball}
(the Laplace/Poisson equations for a ball in $\R^n$). As in
\cite{LA2} the issues of existence and uniqueness are not
addressed; and functions, distributions, boundaries, etc. are
generally assumed to be well-defined in any given case.


\section{Path Integral Solution of PDEs}\label{sec. main}
\subsection{General solution}\label{sec. general}
For reference purposes, the theorem of \cite{LA2} is stated
without proof. Relevant definitions and notation can be found in
\cite{LA2}.
\begin{theorem}\label{main}
Let $\M$ be a real(complex) $m$-dimensional ($m\geq 2$)
paracompact differentiable manifold with a linear connection, and
let $\mathbb{U}$ be a bounded orientable open region in $\M$ with
boundary $\partial{\mathbb{U}}$.\footnote{The boundary
$\p\mathbb{U}$ is assumed to be sufficiently regular.} Let
$\Bold{f}$ and $\Bold\varphi$ be elements of the space of sections
or section distributions of the $(r,s)$-tensor bundle over $\M$.
Assume given the functional $\mathcal{S}(x(\tau_{a'},z))$ whose
associated bilinear form $Q$ satisfies
$\mathrm{Re}(Q(x(\tau_{a'},z))>0$ for $(x(\tau_{a'},z))\neq 0$.

If
$\Bold{\chi}(\x_a\cdot\mathit{\Sigma}(\dual,z))\in\mathcal{F}_R(\Omega)$
where
\begin{eqnarray}\label{chi}
  \Bold{\chi}(\x_a\cdot\mathit{\Sigma}(\dual,z))
  &:=&\int_{C_+}\theta(\dual-\tau_{a'})
  \Bold{f}(\x_a\cdot\mathit{\Sigma}(\tau_{a'},z))
  \exp\left\{-\mathcal{S}(x(\tau_{a'},z))\right\}\,d\tau_{a'}\notag\\
  \notag\\
  &+&\Bold{\varphi}(\x_a\cdot\mathit{\Sigma}(\dual,z))
  \exp\left\{-\mathcal{S}(x(\dual,z))\right\}\;;
\end{eqnarray}
then, for $\x_a=x(\tau_a)\in \mathbb{U}$,
\emph{\begin{equation}\label{inhomogeneous}
  \Bold{\Psi}(\x_a)
  =\int_\Omega
  \Bold{\chi}(\x_a\cdot\mathit{\Sigma}(\taubot,z))\,\DO{\Omega}
  \end{equation}}
 is a solution of the inhomogeneous PDE
\emph{\begin{equation}\label{inhomogeneous PDE}
   \left.\left[\frac{G^{\alpha\beta}}{4\pi}
    \mathcal{L}_{{\Xa}}\mathcal{L}_{{\Xb}}+\mathcal{L}_{\mathbf{Y}}
    +V(\x)\right]\right|_{\x=\x_a}\Bold{\Psi}(\x_a)=-\Bold{f}(\x_a)
\end{equation}}
with boundary condition \emph{\begin{equation}\label{psi limit}
   \Bold{\Psi}(\x_B)
   =\Bold{\varphi}(\x_B)\;.
\end{equation}}
\end{theorem}

\subsection{Elliptic PDEs}\label{sec. elliptic}
Henceforth, restrict to the case where $C_+=\R_+$ or $C_+=i\R$.
Write $\tr=s\widehat{\tau}$ with $s\in\{1,i\}$ and
$\widehat{\tau}\in\R_+$ or $\widehat{\tau}\in\R$ which ever the
case may be. For simplicity I will continue to write $\tau$ in
place of $\widehat{\tau}$ with the understanding that $\tr$ is
real in this context. Although Theorem \ref{main} holds more
generally, it is appropriate to render the construction more
accessible to standard applications and to connect with the
notation of \cite{CA/D-M}, \cite{CA/D-W2}, and \cite{LA}.

Elliptic PDEs are characterized by closed boundaries with
Dirichlet/Neumann boundary conditions and a positive definite $Q$.
It is a simple matter to specialize Theorem \ref{main} for this
case:
\begin{corollary}\label{cor. elliptic}
Given the hypotheses of Theorem \ref{main}, make two
qualifications: Let $\mathbb{U}$ be a bounded orientable region in
$\M$ with \underline{closed} boundary $\partial{\mathbb{U}}$ and
assume the matrix $G^{\alpha\beta}$ has index $(d,0)$. If
$\Bold{\chi}^{(E)}(\x_a\cdot\mathit{\Sigma}(\dual,z))\in\mathcal{F}_R(\Omega)$
where
\begin{eqnarray}
  \Bold{\chi}^{(E)}(\x_a\cdot\mathit{\Sigma}(\dual,z))\notag\\
  &&\hspace{-.4in}:=\int_{C_+}\theta(s[\dual-\tau_{a'}])
  \Bold{f}(\x_a\cdot\mathit{\Sigma}(\tau_{a'},z))
  \exp^{\left\{-s^{-1}\mathcal{S}(x(\tau_{a'},z))\right\}}
  \,d\tau_{a'}\notag\\
  \notag\\
  &&\hspace{-.2in}+\Bold{\phi}(\x_a\cdot\mathit{\Sigma}(\dual,z))
  \exp^{\left\{-s^{-1}\mathcal{S}(x(\dual,z))\right\}}\;;
\end{eqnarray} then, for
$\x_a\in \mathbb{U}$,
\begin{equation}\label{elliptic}
  \Bold{\Psi}^{(E)}(\x_a)
  =\int_\Omega
  \Bold{\chi}^{(E)}(\x_a\cdot\mathit{\Sigma}(\taubot,z))\,\DO{\Omega}\;,
\end{equation}
 is a solution of the inhomogeneous elliptic PDE
\begin{equation}\label{elliptic PDE}
      \left.\left[\frac{s^2}{4\pi}G^{\alpha\beta}
    \mathcal{L}_{{\Xa}}\mathcal{L}_{{\Xb}}+s\mathcal{L}_{\mathbf{Y}}
    +V(\x)\right]\right|_{\x=\x_a}\Bold{\Psi}^{(E)}(\x_a)=-\Bold{f}(\x_a)
   \end{equation}
with boundary condition
\begin{equation}
   \Bold{\Psi}^{(E)}(\x_B)
   =\Bold{\phi}(\x_B)\;.
\end{equation}
\end{corollary}

The corollary clearly follows from Theorem \ref{main}. The only
subtlety is keeping track of factors of $s$. The domain of
integration $C_+$ is now $\R_+$ or $\R$ depending on whether $s=1$
or $s=i$.

The kernels for Dirichlet/Neumann boundary conditions can be
obtained directly from Section $3.2$ of \cite{LA2} and will not be
repeated here. As an example, if the boundary of $\mathbb{U}$ is
at infinity, $V(\x)\rightarrow V(\x)+2\pi\mathcal{E}$ and
$\Bold{\phi}=0$, then (\ref{elliptic}) can be written as the
Fourier/Laplace transform of a path integral solution of an
inhomogeneous parabolic PDE. The kernel of (\ref{elliptic PDE}),
for this case
  with vanishing Dirichlet boundary conditions, is
\begin{equation}\label{elliptic G(D)}
  \K_{\mathbb{U}}^{(D)}(\x_a,\x_{a'};\mathcal{E})
  =\int_{\Za}\int_{C_+}
  \Bold{\delta}(\x_a\cdot\mathit{\Sigma}(\tau_{a'},z),\x_{a'})
  \exp{\left\{-s^{-1}\mathcal{S}(x(\tau_{a'},z);\mathcal{E})\right\}}
  d\tau_{a'}\DQ{z}\;.
\end{equation}
 In quantum physics, (\ref{elliptic
G(D)}) with $s=i$ is the fixed-energy Green's function of the
time-independent inhomogeneous Schr\"{o}dinger equation \emph{when
the boundary is at infinity}. It is the Fourier/Laplace transform
of the position-to-position transition amplitude
$\mathcal{K}(\x_{a'},\tau_{a'};\x_a,\tau_a)$ associated with the
time-dependent Schr\"{o}dinger equation. As previously mentioned,
the Fourier/Laplace transform interpretation becomes a Lagrange
multiplier interpretation for phase space constructions. That is,
(\ref{elliptic G(D)}) can be used (\cite{LA}) to give a phase
space fixed-energy Green's function.

\hspace{.3in}

\emph{Remark}: In \cite{LA}, the integrator $\DO{\tr}$ was chosen
to be Gaussian, which in hindsight was not a good choice. However,
comparing \cite{LA} with the present construction shows that the
theorem in \cite{LA} still holds provided $\DO{\tr}$ is a gamma
integrator. In light of Corollary \ref{cor. elliptic}, the path
integral in the theorem in \cite{LA} solves the inhomogeneous
elliptic PDE with vanishing boundary conditions at infinity.
Consequently, the fixed-energy Green's function calculated in
\cite{LA} is the \emph{elementary kernel with Dirichlet boundary
conditions at infinity} and not the boundary kernel.

\subsection{Parabolic PDEs}\label{sec. parabolic} The path integral
solution to parabolic PDEs when the (open) boundary
$\partial{\mathbb{U}}$ is nowhere tangent to the characteristic
direction has been known for a long time. The solution, in the
general context considered here, was presented in \cite{CA/D-M}.
However, when a segment of $\partial{\mathbb{U}}$ is tangent to
the characteristic direction (for example, a boundary
corresponding to a physical object whose position is fixed in
time), supplemental Dirichlet/Neumann conditions are required
along the segment. Both situations can be handled with some
specialization of Theorem \ref{main}.

\subsubsection{General solution} Because parabolic PDEs lend
themselves to an evolutionary interpretation along the
characteristic direction, it is convenient to assume that $\M$ is
foliated so that the leaves coincide with a real(complex)
$(m-1)$-dimensional paracompact differentiable manifold $\M^-$.
The region of interest will be an orientable submanifold
$\mathbb{U}$ which has the structure $\mathbb{U}^-\times
\mathbb{I}\subseteq \M$ where $\mathbb{U}^-\subseteq \M^-$, and
the interval $\mathbb{I}\subseteq \R_+$ is a subset of the
characteristic manifold of some parabolic PDE. The boundary of
$\M$ is the union of two pieces; the boundary segment
$\partial{\mathbb{U}^-}\times
\mathbb{I}^-:=\partial{\mathbb{U}^-}\times (\mathbb{I}\setminus
\{0\})$ and the Cauchy surface $\mathbb{U}^-\times\{0\}$.

It is necessary to construct a new parametrization for $\M$. Let
$(\x^i,\x^0),\; i\in\{1,\ldots,m-1\}$ denote the coordinates in a
local chart distinguished by the foliation. Instead of fixing the
initial end-point, fix the final end-point; not because it is
necessary but because it is instructive (and it conforms to
physics usage). Hence, the final end-points of paths in $\M^-$ are
fixed according to $x(\tau_{b})=\x_{b}\in \M^-$ and the final
end-points of paths in $\mathbb{I}$ are fixed,
$x^0(\tau_{b})=\x_{b}^0\in \mathbb{I}$.

Let $\Tb$ denote the space of $L^{2,1}$ functions
$\tr:\mathbb{T}=[\ti_a,\ti_b]
\rightarrow[\tau_a,\tau_b]\cup[\tau_a,\tau_b]^*\subset\C_+$ such
that $|d\tr/d\ti|>0$ and $\tau_b:=\tr(\ti_b)=0$. Denote the space
of $L^{2,1}$ pointed paths
$z:\tau(\mathbb{T})\rightarrow\R^d(\C^d)$ with $z(\ti_b)=0$ by
$\Z$. Let $\Omega'=\Z\times\Tb$. Construct a parametrization
$P:\Omega'\rightarrow\PMp$ via the differential equations
\begin{equation}\label{newparametrization}
  \left\{ \begin{array}{ll}
    \Bold{dx}^0(\tr(\ti))-\mathbf{Y}^0(x^0(\tr(\ti)))d\tr=0,\
    & x^0(\tau_{b})=\x^0_{b}\\
    \Bold{dx}(\tr(\ti),z)-\mathbf{Y}(x(\tr(\ti),z))d\tr=
    {\Xa}(x(\tr(\ti),z))\Bold{dz}^\alpha(\tr(\ti)),
    \ & x(\tau_{b})=\x_{b}\\
 \end{array}
  \right.
\end{equation}
where the set of vector fields
$\{\Xa(x(\tr(\ti))),\mathbf{Y}(x(\tr(\ti)))\}\in
T_{x(\tau(\ti))}\M^-$, the vector field
$\mathbf{Y}^0(x(\tr(\ti)))\in T_{x(\tau(\ti))}\mathbb{I}$, and
$[\Xa\,,\,\mathbf{Y}^0]=0$. Evidently there are two independent
sets of parametrized paths; $x(\tau,z)\in \M^-$ and $x^0(\tr)\in
\mathbb{I}$. Since the path $x^0$ will eventually be identified
with the evolution parameter, take
$\mathbf{Y}^0=-\partial/{\partial \x_{b}^0}$ so the
parametrization for $x^0$ according to (\ref{newparametrization})
is just $x^0(\tr)=\x_{b}^0-\tr$.

Since the boundary has two pieces, there will be at least two
critical paths. The critical paths $x_{cr}$ and $x^0_{cr}$ reach
$\p\mathbb{U}$ at the first exit time $\taubotp$. The critical
path that reaches the Cauchy surface satisfies
$x^0_{cr}(\taubotp)=0=\x_{b}^0-\taubotp$ implying
$\taubotp=\x_{b}^0$ for this critical path. Note that
$\taubotp\rightarrow 0$ as $\x_b\rightarrow\x_B$ or
$\x^0_b\rightarrow 0$.

It is now possible to construct the parabolic corollary to Theorem
\ref{main}.
\begin{corollary}\label{cor. parabolic}Assume the same hypotheses
as in Theorem \ref{main}. Endow $\U\subseteq\M$ with the foliated
structure as described above and parametrize according to
Eq.(\ref{newparametrization}). The matrix $G^{\alpha\beta}$ is
assumed to have index $(d,0)$. If
$\Bold{\chi}^{(P)}((\x_{b},\x_{b}^0)\cdot\mathit{\Sigma}(\dual,z))
\in\mathcal{F}_R(\Omega')$ where
\begin{eqnarray}\label{parabolic chi}
  \Bold{\chi}^{(P)}
  ((\x_{b},\x_{b}^0)\cdot\mathit{\Sigma}(\dual,z))\notag\\
  \nonumber\\
  &&\hspace{-1.8in}:=\int_{C_+}\theta(s[\dual-\tau_{a'}])
  \Bold{f}((\x_{b},\x_{b}^0)\cdot\mathit{\Sigma}(\tau_{a'},z))
  \exp^{\left\{-s^{-1}\mathcal{S}((x,x^0)(\tau_{a'},z))\right\}}
  \,d\tau_{a'}\nonumber\\\notag\\
  &&\hspace{-1.5in}+
  \Bold{\varphi}((\x_{b},\x_{b}^0)\cdot\mathit{\Sigma}(\dual,z))
  \exp^{\left\{-s^{-1}\mathcal{S}((x,x^0)(\dual,z))\right\}}
  \;,\notag\\
\end{eqnarray}
then, for $(\x_{b},\x_{b}^0)\in \mathbb{U}\times \mathbb{I}$,
\begin{equation}\label{parabolic integral}
  \Bold{\Psi}^{(P)}(\x_{b},\x_{b}^0)=\int_{\Omega'}
  \Bold{\chi}^{(P)}((\x_{b},\x_{b}^0)\cdot\mathit{\Sigma}(\taubotp,z))
  \,\DO{\Omega'}\;,
\end{equation}
is a solution of the inhomogeneous parabolic PDE
\begin{eqnarray}\label{inhomogeneous parabolic}
 \left.\left[\frac{s^2}{4\pi}G^{\alpha\beta}
    \mathcal{L}_{{\Xa}}\mathcal{L}_{{\Xb}}+s\mathcal{L}_{\mathbf{Y}}
    +V(\x)
    -s\frac{\partial}{\partial{\x^0}}\right]
    \right|_{\x=(\x_{b},\x_{b}^0)}
    \Bold{\Psi}^{(P)}(\x_{b},\x_{b}^0)\nonumber\\\nonumber\\
    =-\Bold{f}(\x_{b},\x_{b}^0)&&
\end{eqnarray}
with initial and boundary conditions
\begin{equation}\label{BC}
\left.\Bold{\Psi}^{(P)}(\x_{b},\x_{b}^0)\right|_{\x^0_{b}=0}
=\Bold{\varphi}(\x_{b},0)\;,\ \ \ \
\left.\Bold{\Psi}^{(P)}(\x_{b},\x_{b}^0)\right|_{\x_{b}=\x_B}
=\Bold{\varphi}(\x_B,\x_{b}^0)\;.
\end{equation}
\end{corollary}

\emph{Proof.} It is not difficult to check that this corollary is
a consequence of Theorem \ref{main}. The only substantive
difference is the presence of the $\p/\p\x_{b}^0$ term in the
partial differential operator, which is a consequence of the
parametrization (\ref{newparametrization}) and the choice
$\mathbf{Y}^0\propto\partial/{\partial \x_{b}^0}$. To show that
the initial and boundary conditions follow from the theorem, it
helps to use $\partial{\mathbb{U}}=(\partial{\mathbb{U}^-}\times
\mathbb{I}^-)\cup (\mathbb{U}^-\times\{0\})$ in condition
(\ref{psi limit}), and to remember that $\taubotp\rightarrow 0$ as
the boundary is approached.$\QED$

As a bonus, a ``time-dependent'' potential $V(\x_{a'},\x_{a'}^0)$
has been achieved. Also note that (\ref{inhomogeneous parabolic})
is a diffusion-type equation for $s$ real and a
Schr\"{o}dinger-type equation for $s$ imaginary.

\subsubsection{Elementary kernels}\label{parabolic kernels}

\begin{lemma}
The parabolic Dirichlet elementary kernel is given by
\begin{eqnarray}\label{interior K(D)}
  \K_\mathbb{U}^{(D)}((\x_{b},\x_{b}^0),(\x_a,\x^0_a))\!\!\!\!
  &:=&\!\!\!\!\int_{\Omega'}\taurange
 \theta(s[\taubotp-\tau_{a'}])
  \Bold{\delta}(\x_{b}\cdot\mathit{\Sigma}(\tau_{a'},z),\x_a)
  \delta((\x_b^0-\x_{a}^0)-\tau_{a'})\nonumber \\
  &&\hspace{.4in}
  \times\exp\left\{-s^{-1}\mathcal{S}((x,x^0)(\tau_{a'},z))\right\}
  \;d\tau_{a'}\DO{\Omega'}\;.
\end{eqnarray}
\end{lemma}

\emph{Proof.} As in the elliptic case, the elementary kernel for
the parabolic PDE is obtained by fixing the end-point with the
delta functional
\begin{equation}
  \Bold{\delta}((\x_{b},\x_{b}^0)
  \cdot\mathit{\Sigma}(\tau_{a'},z),(\x_a,\x_a^0))\;.
\end{equation}
From the parametrization for $x^0$ follows
$x^0(\tau_{a'})=\x_{b}^0-\tau_{a'}$ which implies the delta
functional reduces to
\begin{equation}
  \Bold{\delta}(\x_{b}\cdot\mathit{\Sigma}(\tau_{a'},z),\x_a)
  \delta((\x_b^0-\tau_{a'})-\x_{a}^0)
\end{equation}
with $|\x_b^0-\x_{a}^0|\geq 0$. It is straightforward to check
that the integral has the correct form to apply the theorem, and
that the kernel vanishes on the boundary. $\QED$

It is useful to separate the boundary kernel into a kernel
$\K_{\partial{\mathbb{U}^-}}^{(D)}$ on the boundary segment
$\partial{\mathbb{U}^-}\times \mathbb{I}^-$ and a kernel
$\K_C^{(D)}$ on the Cauchy surface $\mathbb{U}^-\times \{0\}$.

\begin{lemma}
The parabolic Dirichlet boundary segment kernel is given by
\begin{eqnarray}\label{boundary K(D)}
 \K_{\partial{\mathbb{U}^-}}^{(D)}((\x_{b},\x_{b}^0),(\x_B,\x_B^0))
  &:=&\int_{\Omega'}
  \Bold{\delta}(\x_{b}\cdot\mathit{\Sigma}(\taubotp,z),\x_B)
  \delta((\x_b^0-\x_{B}^0)-\taubot)
  \nonumber \\
  &&\hspace{.2in}\times\exp\left\{-s^{-1}\mathcal{S}((x,x^0)(\taubotp,z))
  \right\}
  \;\DO{\Omega'}\;,
  \end{eqnarray}
and the parabolic Dirichlet Cauchy kernel by
\begin{eqnarray}\label{second homogeneous K(D)}
  \K_{C}^{(D)}((\x_{b},\x_{b}^0),(\x_a,0))
  &:=&\int_{\Z}
  \Bold{\delta}(\x_{b}\cdot\mathit{\Sigma}(\x_{b}^0,z),\x_a)
  \nonumber\\
  &&\hspace{.2in}
  \times\exp\left\{-s^{-1}\mathcal{S}((x,x^0)(\x_{b}^0,z))\right\}
  \;\DQ{z}\;.\
\end{eqnarray}
\end{lemma}

\emph{Proof.} The boundary segment kernel clearly satisfies the
homogeneous parabolic PDE and possesses the correct boundary
condition. In specifying the Cauchy kernel, be mindful that the
Cauchy kernel corresponds to a transition from a point $\x_a$ on
the Cauchy surface ($\x_a^0=0$) to some point $\x_{b}$ at the time
$\x_{b}^0$---or vice versa:
  \begin{eqnarray}\label{Cauchy K(D)}
  \K_{C}^{(D)}((\x_{b},\x_{b}^0),(\x_a,0))
  &:=&\int_{\Omega'}
  \Bold{\delta}(\x_{b}\cdot\mathit{\Sigma}(\taubotp,z),\x_a)
   \delta(\x_{b}^0-\taubotp)
  \nonumber\\
  &&\hspace{.2in}
  \times\exp\left\{-s^{-1}\mathcal{S}((x,x^0)(\taubotp,z))\right\}
  \;\DO{\Omega'}
  \nonumber\\
  &=&\mathcal{N}\int_{\Z}\taurange
  \Bold{\delta}(\x_{b}\cdot\mathit{\Sigma}(\taubotp,z),\x_a)
   \delta(\x_{b}^0-\taubotp)
  \nonumber\\
  &&\hspace{.2in}
  \times\exp\left\{-s^{-1}\mathcal{S}((x,x^0)(\taubotp,z))\right\}
  \;d(\ln\taubotp)\,\DQ{z}
  \nonumber\\
  &=&\int_{\Z}
  \Bold{\delta}(\x_{b}\cdot\mathit{\Sigma}(\x_{b}^0,z),\x_a)
  \nonumber\\
  &&\hspace{.2in}
  \times\exp\left\{-s^{-1}\mathcal{S}((x,x^0)(\x_{b}^0,z))\right\}
  \;\DQ{z}\;.\
\end{eqnarray}
In the third equality, the constant $\mathcal{N}$ has been chosen
to give $\K_{C}^{(D)}((\x_{b},\x_{b}^0),(\x_a,0))$ the correct
normalization. $\QED$

This is the well-known path integral kernel for parabolic PDEs for
unbounded manifolds. For consistency, the Cauchy kernel must agree
with the elementary kernel evaluated at $\x_a^0=0$ for the
boundary at infinity. Putting $\x_a^0=0$ and
$\taubotp\rightarrow\infty$ in (\ref{interior K(D)}) indeed yields
(\ref{second homogeneous K(D)}).

\begin{corollary} The solution to the inhomogeneous parabolic PDE with
Dirichlet conditions $\Bold{\phi}(\x_B,\x_{b}^0)$ and Cauchy
conditions $\Bold{\psi}(\x_{b},0)$ is
\begin{eqnarray}\label{parabolic Dirichlet kernel solution}
  \Bold{\Psi}^{(P)}(\x_{b},\x_{b}^0)
  &=&\int_{\mathbb{U}^-}\int_{\Omega'}
 \theta(s[\taubotp-(\x_{b}^0-\x_a^0)])
  \Bold{\delta}(\x_{b}\cdot\mathit{\Sigma}((\x_{b}^0-\x_a^0),z),\x_a)
  \Bold{f}(\x_a,\x_a^0)
  \nonumber \\
  &&\hspace{.7in}
  \times\exp\left\{-s^{-1}\mathcal{S}((x,x^0)((\x_{b}^0-\x_a^0),z))
  \right\}
  \;\DO{\Omega'}\,d\x_a\nonumber \\
  &&+\int_{\partial{\mathbb{U}^-}}\int_{\Omega'}
  \Bold{\delta}(\x_{b}\cdot\mathit{\Sigma}(\taubotp,z),\x_{B'})
      \Bold{\phi}(\x_{B'},(\x^0_{b}-\taubotp))\nonumber \\
  &&\hspace{.7in}
  \times\exp\left\{-s^{-1}\mathcal{S}((x,x^0)(\taubotp,z))\right\}
  \;\DO{\Omega'}\,d\x_{B'} \nonumber \\
  &&+\int_{\mathbb{U}^-}\int_{\Z}\Bold{\delta}(\x_{b}
  \cdot\mathit{\Sigma}(\x_{b}^0,z),\x_a)
    \Bold{\psi}(\x_a,0) \nonumber\\
  &&\hspace{.7in}
  \times\exp\left\{-s^{-1}\mathcal{S}((x,x^0)(\x_{b}^0,z))\right\}
  \;\DQ{z}\,d\x_a\notag\\\notag\\
  &=&\int_{\Omega'}
 \theta(s[\taubotp-(\x_{b}^0-\x_a^0)])
 \Bold{f}(\x_{b}\cdot\mathit{\Sigma}((\x_{b}^0-\x_a^0),z),\x_a^0)
  \nonumber \\
  &&\hspace{.7in}
  \times\exp\left\{-s^{-1}\mathcal{S}((x,x^0)((\x_{b}^0-\x_a^0),z))
  \right\}
  \,\DO{\Omega'}\nonumber \\
  &&+\int_{\Omega'}
 \Bold{\phi}(\x_{b}\cdot\mathit{\Sigma}(\taubotp,z),(\x^0_{b}-\taubotp))
 \nonumber \\
  &&\hspace{.7in}
  \times\exp\left\{-s^{-1}\mathcal{S}((x,x^0)(\taubotp,z))\right\}
  \,\DO{\Omega'}\nonumber \\
  &&+\int_{\Z}
    \Bold{\psi}(\x_{b}\cdot\mathit{\Sigma}(\x_{b}^0,z),0) \nonumber\\
  &&\hspace{.7in}
  \times\exp\left\{-s^{-1}\mathcal{S}((x,x^0)(\x_{b}^0,z))\right\}
  \;\DQ{z}\,d\x_a\;.
\end{eqnarray}
\end{corollary}

\emph{Proof.} The expression follows from the two preceding
lemmas. To check the boundary conditions, the first term vanishes
for both $\x_{b}\rightarrow \x_{B}$ and $\x_{b}^0\rightarrow 0$
since $\taubotp\rightarrow 0$ in both cases. For the limit
$\x_{b}\rightarrow \x_{B}$, the second term gives the boundary
condition $\Bold{\phi}(\x_{B},\x_{b}^0)$ along
$\partial{\mathbb{U}^-}\times \mathbb{I}^-$. The third term
contributes $\Bold{\psi}(\x_B,0)=\Bold{\phi}(\x_B,0)$ on
$\partial{\mathbb{U}^-}\times \{0\}$. (In the third term,
$\x_b\rightarrow\x_B$ implies $\x_b^0\rightarrow 0$. This follows
from the defining equation of
$\K_{C}^{(D)}((\x_{b},\x_{b}^0),(\x_a,0))$ in (\ref{Cauchy
K(D)}).) When $\x_{b}^0\rightarrow 0$, the second term gives no
contribution because $\taubotp\nless 0$. (It must be kept in mind
that $(x_b^0-\taubotp)>0$ in the second term, because the
integration over $\mathbb{I}^-$ did not include the point
$\{0\}\in\mathbb{I}$.) The third term gives the initial condition
$\Bold{\psi}(\x_{a'},0)$. $\QED$

Observe that (\ref{parabolic  Dirichlet kernel solution})
satisfies the homogeneous PDE with Cauchy initial conditions only
(with vanishing boundary conditions at infinity) by the particular
choice $\Bold{f}(x)=0$ and $\Bold{\phi}(x)=0$. Specifically,
\begin{eqnarray}\label{parabolic path integral}
   \Bold{\Psi}^{(P)}(\x_{b},\x_{b}^0)
   &=&\int_{\Z}\Bold{\psi}(\x_{b}\cdot\mathit{\Sigma}(\x_{b}^0,z))
  \exp\left\{-s^{-1}\mathcal{S}((x,x^0)(\x_{b}^0,z))\right\}\;\DQ{z}
  \;.\nonumber\\
\end{eqnarray}
With the identification $\x_{b}^0\equiv \ti_b$, this is the
solution presented in \cite{CA/D-M} for the parabolic PDE with
vanishing boundary conditions at infinity (as it should be).

The kernels for the case of Neumann boundary conditions for
non-compact $\M$ can be adapted from Subsection 3.2.2 in
\cite{LA2}.
\begin{lemma}
The parabolic Neumann elementary kernel is given by
\begin{eqnarray}
  \K_\mathbb{U}^{(N)}((\x_{b},\x_{b}^0),(\x_a,\x_a^0))
  &:=&\K_{\infty}((\x_{b},\x_{b}^0),(\x_a,\x_a^0))
  +\mathbf{F}_\mathbb{U}((\x_{b},\x_{b}^0),(\x_a,\x_a^0))\nonumber\\
\end{eqnarray}
where $\K_{\infty}$ and $\mathbf{F}_\mathbb{U}$ are now defined in
an obvious way from (\ref{interior K(D)}).
\end{lemma}

\begin{lemma}
The parabolic Neumann boundary segment kernel is given by
\begin{eqnarray}\label{K(N) boundary}
  \K^{(N)}_{\partial \mathbb{U}^-}((\x_{b},\x_{b}^0),(\x_B,\x_B^0))
  &:=&\int_{\Omega'}
  \Bold{\theta}(x(\taubot,z),\x_B)\delta((\x_{b}^0-\x_B^0)-\taubotp)
  \nonumber\\
 &&\hspace{.2in}\times
  \exp{\left\{-s^{-1}\mathcal{S}((x,x^0)(\taubotp,z))\right\}}
  \;\DO{\Omega'}\;,
\end{eqnarray}
and the parabolic Neumann Cauchy kernel by
\begin{eqnarray}\label{Neumann Cauchy}
\K^{(N)}_{C}((\x_{b},\x_{b}^0),(\x_a,0))
  &:=&-\int_{\Omega'}
  \Bold{\delta}(\x_{b}\cdot\mathit{\Sigma}(\taubotp,z),\x_a)
  \theta(\x_{b}^0-\taubotp)
 \nonumber\\
  &&\hspace{.4in}\times
  \exp\left\{-s^{-1}\mathcal{S}((x,x^0)(\taubotp,z))\right\}
  \,\DO{\Omega'}\;.
\end{eqnarray}
\end{lemma}

The proofs of these two lemmas will be omitted, because they are
straightforward specializations of the general Neumann kernels in
section 3.2.2 of \cite{LA2} and they essentially repeat previous
arguments.

\begin{corollary}
The solution to the inhomogeneous parabolic PDE with Neumann
conditions
$\nabla_{\mathbf{n}_{\partial}}\Bold{\phi}(\x_{B},\x_{b}^0)$ and
Cauchy conditions
$\nabla_{\partial/\partial{\x_{b}^0}}\Bold{\psi}(\x_{b},0)$ is
\begin{eqnarray}
\Bold{\Psi}^{(P)}(\x_{b},\x_{b}^0)
  &=&\int_{\mathbb{U}^-}\int_{\Omega'}
  \K_\mathbb{U}^{(N)}((\x_{b},\x_{b}^0),(\x_a,\x_a^0))
  \Bold{f}(\x_a,\x_a^0)
  \nonumber \\
  &&\hspace{.7in}
  \times\exp\left\{-s^{-1}\mathcal{S}((x,x^0)((\x_{b}^0-\x_a^0),z))
  \right\}
  \,\DO{\Omega'}\,d\x_a\nonumber \\
  &&+\int_{\partial{\mathbb{U}^-}}\int_{\Omega'}
  \Bold{\theta}(x(\taubotp,z),\x_{B'})
  \nabla'_{\mathbf{n}_{\partial}}\Bold{\phi}(\x_{B'},(\x_{b}^0-\taubotp))
  \nonumber\\
  &&\hspace{1.5in}
  \times\exp\left\{-s^{-1}\mathcal{S}((x,x^0)(\taubotp,z))\right\}
  \,\DO{\Omega'}\,d\x_{B'}\nonumber \\
  &&-\int_{\Omega'}\theta(\x_{b}^0-\taubotp)
 \nabla_{\partial/\partial{\x_{b}^0}}
  \Bold{\psi}(\x_{b}\cdot\mathit{\Sigma}(\taubotp,z),0)\nonumber\\
  &&\hspace{.5in}\times
  \exp\left\{-s^{-1}\mathcal{S}((x,x^0)(\taubotp,z))\right\}
  \,\DO{\Omega'}\;.\notag\\
\end{eqnarray}

\end{corollary}

\emph{Proof.} The terms follow from the preceding lemmas. It
remains to check the boundary conditions for
$\nabla_{\mathbf{n}_{\partial}}\Bold{\Psi}^{(P)}$.  When
$\x_{b}^0\rightarrow 0$, $\mathbf{n}_{\partial}$ is in the
$\x_{b}^0$ direction. Neither $ \K_\mathbb{U}^{(N)}$ nor
$\K^{(N)}_{\partial \mathbb{U}^-}$(since $\taubotp\nless 0$)
contribute. The third term gives the Cauchy initial condition
$\nabla_{\partial/\partial{\x_{b}^0}}\Bold{\psi}(\x_{b},0)$,
because the derivative on the $\theta$ factor gives a delta
function and the remaining $\theta$ terms vanish for
$\x_{b}^0\rightarrow 0$. When $\x_{b}\rightarrow \x_B$,
$\nabla_{\mathbf{n}_{\partial}}\Bold{\Psi}^{(P)}$ gets no
contribution from the first term and the second term gives the
Neumann conditions
$\nabla_{\mathbf{n}_{\partial}}\Bold{\phi}(\x_B,\x_{b}^0)$ for
$\x_b^0>0$ due to the derivative on the $\Bold{\theta}$ factor and
the vanishing of the remaining $\Bold{\theta}$ terms on the
boundary. The third term contributes the Neumann conditions on the
Cauchy surface $\nabla_{\mathbf{n}_{\partial}}\Bold{\psi}(\x_B,0)
=\nabla_{\mathbf{n}_{\partial}}\Bold{\phi}(\x_B,0)$ after
recalling that $\x_b\rightarrow\x_B$ implies $\x_b^0\rightarrow 0$
(in the third term) and reducing the integral over $\Tb$ to a
one-dimensional integral and integrating by parts. $\QED$

Note that it is possible to mix Dirichlet and Neumann conditions
on the boundary segment and Cauchy surface. The expression for the
associated $\Bold{\Psi}^{(P)}$ will have the obvious collection of
relevant kernels.

\subsection{Hyperbolic PDEs}\label{sec. hyperbolic} Like the
elliptic and parabolic cases, hyperbolic PDEs can be solved by
specializing Theorem \ref{main}. In fact, the solution is a rather
trivial modification of Corollary \ref{cor. parabolic}.

Assume that the matrix $G^{\alpha\beta}$ has index $(d-r,r)$ for
some $r\in\{1,\ldots,d\}$. The quadratic form $Q(x(\tau_{a'},z))$
will have the same index since $G^{\alpha\beta}$ is ultimately
determined by the inverse of $Q$. Recall that $\mathrm{Re}(Q)$
must be positive definite. It follows that the hyperbolic case
will be associated with an imaginary $Q$. Furthermore, at least
for wave equations, it is known that the boundary of interest must
be open. (Whether this holds in the general case is, again, an
open question.) With these qualifications, Theorem \ref{main} can
be applied directly.

Because of its importance, it is useful to spell out the details
for the case $r=1$. It is convenient to formulate the corollary in
terms of a foliated $\M$ as in Section \ref{sec. parabolic} and to
fix the final end-points. The leaves of the foliation are assumed
to be space-like. The evolution parameter is no longer along a
characteristic direction, and the parametrization is determined by
\begin{equation}\label{hyper param}
  \left\{ \begin{array}{ll}
    \Bold{dx}^0(\tr(\ti),z^0)-\mathbf{Y}^0(x^0(\tr(\ti),z^0))d\tr
    =\mathbf{X}^0(x^0(\tr(\ti),z^0))\Bold{dz}^0(\tr(\ti))
    ,\ & x^0(\tau_{b})=\x^0_{b}\\
    \Bold{dx}(\tr(\ti),z)-\mathbf{Y}(x(\tr(\ti),z))d\tr=
    {\Xa}(x(\tr(\ti),z))\Bold{dz}^\alpha(\tr(\ti)),
    \ & x(\tau_{b})=\x_{b}\\
 \end{array}
  \right.
\end{equation}
where the space of paths is $\Z$.
\begin{corollary}\label{cor-hyperbolic}
Assume the same hypotheses as Corollary \ref{cor. parabolic} with
the following qualifications: The parametrization is determined by
Eq. (\ref{hyper param}), the matrix $G^{\alpha\beta}$ has index
$(d-1,1)$, and $\mathrm{Re}(Q(x(\tau_{a'},z)))=0$. If
$\Bold{\chi}^{(H)}((\x_{b},\x_{b}^0)\cdot\mathit{\Sigma}(\dual,z))
\in\mathcal{F}_R(\Omega')$ with
$\Bold{\chi}^{(H)}((\x_{b},\x_{b}^0)\cdot\mathit{\Sigma}(\dual,z))$
given by (\ref{parabolic chi}), then
\begin{equation}
  \Bold{\Psi}^{(H)}(\x_{b},\x_{b}^0)=\int_{\Omega'}
  \Bold{\chi}^{(H)}((\x_{b},\x_{b}^0)\cdot\mathit{\Sigma}(\taubotp,z))
  \,\DO{\Omega'}\;,
\end{equation} is a solution of the
inhomogeneous wave equation
\begin{equation}\label{homogeneous hyperbolic}
    \left.\left[\frac{s^2}{4\pi}G^{\alpha\beta}
    \mathcal{L}_{{\Xa}}\mathcal{L}_{{\Xb}}+s\mathcal{L}_{\mathbf{Y}}
    +V(\x)\right]\right|_{\x=(\x_{b},\x_{b}^0)}
    \Bold{\Psi}^{(H)}(\x_{b},\x_{b}^0)
    =-\Bold{f}(\x_{b},\x_{b}^0)
\end{equation}
with initial conditions
\begin{eqnarray}
  \left.\Bold{\Psi}^{(H)}(\x_{b},\x_{b}^0)
  \right|_{\x^0_{b}=0}
  &=&\Bold{\varphi}(\x_{b},0)\,,\nonumber\\
   \left.\nabla_{\partial/\partial{\x_{b}^0}}
   \Bold{\Psi}^{(H)}(\x_{b},\x_{b}^0)\right|_{\x^0_{b}=0}
   &=&\nabla_{\partial/\partial{\x_{b}^0}}\Bold{\varphi}(\x_{b},0)\,,
\end{eqnarray}
and boundary conditions
\begin{equation}
   \left.\Bold{\Psi}^{(H)}(\x_{b},\x_{b}^0)\right|_{\x_{b}=\x_B}
   =\Bold{\varphi}(\x_B,\x_{b}^0)\;.
\end{equation}

\end{corollary}

The kernels can be taken over from the parabolic case remembering
that
\begin{equation}
  \Bold{\delta}((\x_{b},\x_{b}^0)
  \cdot\mathit{\Sigma}(\taubotp,z),(\x_a,\x_a^0))
  \neq\Bold{\delta}(\x_{b}
  \cdot\mathit{\Sigma}(\taubotp,z),\x_a)
  \delta((\x_{b}^0-\taubotp)-\x_a^0)
\end{equation}
in general for the parametrization (\ref{hyper param}). So the
Dirichlet kernels for the wave equation are just the parabolic
Dirichlet kernels with
\begin{equation}
\Bold{\delta}(\x_{b}
  \cdot\mathit{\Sigma}(\taubotp,z),\x_a)
  \delta((\x_{b}^0-\taubotp)-\x_a^0)\longrightarrow
  \Bold{\delta}((\x_{b},\x_{b}^0)
  \cdot\mathit{\Sigma}(\taubotp,z),(\x_a,\x_a^0))\;.
\end{equation}
The Neumann Cauchy kernel is obtained by the substitution
\begin{equation}
 \Bold{\delta}(\x_{b}\cdot\mathit{\Sigma}(\taubotp,z),\x_a)
  \theta(\x_{b}^0-\taubotp)\longrightarrow
\Bold{\delta}(\x_{b}\cdot\mathit{\Sigma}(\taubotp,z),\x_a)
\theta(x(\taubotp),0)\;.
\end{equation}
where
\begin{equation}
\theta(\x_2,\x_1):=\left\{\begin{array}{ll}
0\hspace{.5in}\mbox{for}\hspace{.5in}\x_1\preceq\x_2\\
1\hspace{.5in}\mbox{for}\hspace{.5in}\x_1\succ \x_2
\end{array}\right.
\end{equation}
and the ordering is with respect to the foliation.

The only task is to verify the correct boundary conditions.

\begin{corollary} The solution to the inhomogeneous wave equation
with Cauchy data $\Bold{\psi}(\x_{b},0)$ and
$\nabla_{\partial/\partial{\x_{b}^0}}\Bold{\psi}(\x_{b},0)$ and
Dirichlet boundary conditions $\Bold{\phi}(\x_B,\x_{b}^0)$ is
\begin{eqnarray}\label{wave kernels}
  \Bold{\Psi}^{(H)}(\x_{b},\x_{b}^0)
  &=&\int_\mathbb{U}\int_\mathbb{I}\K_\mathbb{U}^{(D)}
  ((\x_{b},\x_{b}^0),(\x_a,\x_a^0))
  \Bold{f}(\x_a,\x_a^0)
  \,d\x_a\,d\x_a^0\nonumber \\
  &&+\int_{\partial{\mathbb{U}^-}\times
  \mathbb{I}^-}\hspace{-.1in}\K_{\partial \mathbb{U}^-}^{(D)}
  ((\x_{b},\x_{b}^0),(\x_{B'},\x_{B'}^0))
  \Bold{\phi}(\x_{B'},\x_{B'}^0)\;d\x_{B'}\,d\x_{B'}^0 \nonumber \\
  &&+\frac{1}{2}\int_{\mathbb{U}^-}\hspace{-.1in}\K_C^{(D)}
  ((\x_{b},\x_{b}^0),(\x_a,0))
  \Bold{\psi}(\x_a,0)\;d\x_a \nonumber\\
  &&-\frac{1}{2}\int_{\mathbb{U}^-}\hspace{-.1in}
  \K^{(N)}_C((\x_{b},\x_{b}^0),(\x_a,0))
  \nabla_{\partial/\partial{\x_{b}^0}}
  \Bold{\psi}(\x_a,0)\;d\x_a\;.\nonumber\\
\end{eqnarray}
\end{corollary}

\emph{Proof.} To show the boundary conditions are satisfied,
consider the limit $\x_{b}^0\rightarrow 0$ in $\Bold{\Psi}^{(H)}$.
The first and second terms vanish (recall that
$0\notin\mathbb{I}^-$). The third term gives
$1/2\Bold{\psi}(\x_{b},0)$. In this limit, the fourth term can be
integrated by parts (after reducing the $\Tb$ integral as before)
to yield the remaining $1/2\Bold{\psi}(\x_{b},0)$. The same limit
in $\nabla_{\partial/\partial{\x_{b}^0}}\Bold{\Psi}^{(H)}$ gets no
contribution from the first two terms for the same reasons. The
third term gives
$1/2\nabla_{\partial/\partial{\x_{b}^0}}\Bold{\psi}(\x_{b},0)$
after reducing the $\Tb$ integral and an integration by parts. The
fourth term contributes the remaining
$1/2\nabla_{\partial/\partial{\x_{b}^0}}\Bold{\psi}(\x_{b},0)$.
When $\x_{b}\rightarrow \x_B$, the first term vanishes and the
second term yields $\Bold{\phi}(\x_B,\x_{b}^0)$. The third and
fourth terms give
$1/2\Bold{\psi}(\x_B,0)+1/2\Bold{\psi}(\x_B,0)=\Bold{\phi}(\x_B,0)$
as in previous cases. The fourth term requires an integration by
parts as before.$\QED$

Neumann boundary conditions can be handled using the results of
Subsection \ref{parabolic kernels}. The details will be omitted
since they just repeat previous arguments.

\section{Examples}\label{sec. examples}
As it stands, Theorem \ref{main} should be characterized as a
representation of a PDE solution rather than as a prescription for
calculating the solution. The reason is that the complications
introduced into the problem due to the geometry are encoded in
$\taubot$, and one anticipates that techniques will have to be
developed to handle these complications. Indeed, the parameter
$\taubot$ has been defined only implicitly.

In the following examples, three such techniques are introduced in
order to calculate the kernels for some selected geometries. The
first is simply the observation that $\taubot$ and $\DO{\tr}$ play
no role when the boundary is at infinity. The second technique
relies on the result of Appendix \ref{app. universal covering} to
express the kernel of a bounded region in terms of the kernel on
its covering space. The third technique is based on Subsection
3.2.1 in \cite{LA2} and is akin to the method of images.

In all of the examples,  I will work in $\R^n$ with Cartesian
coordinates for simplicity. Since the target manifold is $\R^n$,
it is natural to use the exponential map parametrization which is
explained in Appendix \ref{app. exp}.\footnote{Of course, the
final expressions do not depend on the particular parametrization
that one chooses.} Furthermore, I will take $\mathcal{S}(x)$ with
$V(x)=0$. Well known expansion techniques can be used for general
$V(x)$.

\subsection{Boundary at Infinity}\label{sec. boundary at infinity}
The easiest case to handle is when the boundary (excluding a
possible Cauchy surface) is at infinity. Since the boundary is at
infinity, $\taubot\rightarrow\infty$ (for any physically
reasonable system at least).

\subsubsection{Elliptic Case}\label{elliptic case}
For the elliptic case, take vanishing Dirichlet boundary
conditions at infinity. Only the elementary kernel
$\K_\mathbb{U}^{(D)}$ is needed in this case. Use the exponential
map parametrization for a classical path with $\mathbf{Y}=0$ and
$\mathbf{X}_{\alpha}=\delta_{\alpha}^i\frac{\partial}{\partial
\x^i}$ where $i\in\{1,\ldots,n\}$ and $\alpha\in\{1,\ldots,n\}$.

The Dirichlet elementary kernel is given by
\begin{equation}
  \K_\mathbb{U}^{(D)}(\x_a,\x_{a'})=\int_{\Omega}\taurange
  \Bold{\delta}(x(\tau_{a'},z)-\x_{a'})
  \exp{\left\{-s^{-1}\mathcal{S}(x(\tau_{a'},z))\right\}}
  \;d\tau_{a'}\,\DO{\Omega}\;.
\end{equation}
The parametrization $P$ yields
\begin{equation}
  x(\tr,z)=\x_a+z(\tr)\;.
\end{equation}
The associated quadratic form is
\begin{eqnarray}
  Q(x(\tau_{a'},z))
  &=&\int_0^{\tau_{a'}}
  \delta_{ij}\dot{\Bold{x}}^{i}(\tau,z)\dot{\Bold{x}}^{j}(\tau,z)d\tau
  \nonumber\\
  &=&\int_0^{\tau_{a'}}
  \delta_{\alpha\beta}\dot{\Bold{z}}^{\alpha}(\tau)
  \dot{\Bold{z}}^{\beta}(\tau)d\tau
\end{eqnarray}
so that $G^{\alpha\beta}=\delta^{\alpha\beta}$. Consequently,
$\K_\mathbb{U}^{(D)}(\x_a,\x_{a'})$ satisfies
\begin{equation}
  \frac{s^2}{4\pi}\Delta\K_\mathbb{U}^{(D)}(\x_a\,x_{a'})
  =-\Bold{\delta}(\x_a-\x_{a'})
\end{equation}
where $\Delta$ is the Laplacian on $\R^n$.

On the other hand, the exponential map parametrization gives
\begin{equation}
 x(\tr,\Bold{\zeta})=x_{cr}(\tr)+\Bold{\zeta}(\tr)
\end{equation}
where $x_{cr}$ is a critical point of $\mathcal{S}$ relative to a
variation with both end-points fixed, viz.
\begin{equation}\label{critical path}
  x_{cr}(\tr)=\x_a+\left(\frac{\x_{a'}-\x_a}{\tau_{a'}}\right)\tr\;.
\end{equation}
(Since the boundary is at infinity, this represents a transition
between interior points so the relevant variational problem is for
point-to-point.)

The Gaussian integrator is invariant under translations, and so
the integral can be expressed in terms of $\Bold{\zeta}$ instead:
\begin{equation}
  \K_\mathbb{U}^{(D)}(\x_a,\x_{a'})=\int_{\widetilde{\Omega}}\taurange
  \Bold{\delta}(x(\tau_{a'},\Bold{\zeta})-\x_{a'})
  \exp{\left\{-s^{-1}\pi \widetilde{Q}\right\}}\;d\tau_{a'}
  \,\DO{\widetilde{\Omega}}\;,
\end{equation}
where\footnote{There is no cross term since $\Bold{\zeta}(0)=0$
for the exponential map parametrization, and
$\Bold{\zeta}(\tau_{a'})=0$ will not contribute by virtue of
(\ref{delta}).}
\begin{equation}
 \widetilde{Q}(x(\tau_{a'},{\Bold{\zeta}}))
   =\frac{|\x_{a'}-\x_a|^2}{\tau_{a'}}+\int_0^{\tau_{a'}}
  \delta_{\alpha\beta}
  \dot{{\Bold{\zeta}}}^{\alpha}(\tau)
  \dot{{\Bold{\zeta}}}^{\beta}(\tau)d\tau\;.
\end{equation}

To evaluate $\K_\mathbb{U}^{(D)}$, use\footnote{Recall from
\cite{LA2} that $\Bold{\delta}(\x_a,\x_{a'})$ denotes a Dirac
bitensor composed of Kronecker delta symbols (which are
collectively denoted by $\mathbf{1}$ for brevity) and the scalar
Dirac delta function $\delta(\x_a,\x_{a'})$.}
\begin{eqnarray}\label{delta}
  \Bold{\delta}(x(\tau_{a'},\Bold{\zeta})-\x_{a'})
  &=&\Bold{\delta}(x_{cr}(\tau_{a'})+\Bold{\zeta}(\tau_{a'})-\x_{a'})
  =\int_{\R^n}\mathbf{1}\exp^{\{-2\pi i(\Bold{\mathrm{u}}\cdot
  \Bold{\zeta}(\tau_{a'}))\}}\,d\Bold{\mathrm{u}} \nonumber\\
  &=&\int_{\R^n}\mathbf{1}
  \exp\{-2\pi i\langle \Bold{\mathrm{u}}
  \delta_{\Bold{\zeta}(\tau_{a'})},\Bold{\zeta}\rangle\}
  \,d\Bold{\mathrm{u}}
\end{eqnarray}
to get
\begin{eqnarray}
\K_\mathbb{U}^{(D)}(\x_a,\x_{a'})&=&\int_{\widetilde{\Omega}}
\taurange\int_{\R^n} \mathbf{1}\exp{\left\{-s^{-1}\pi
\widetilde{Q}-2\pi i(\langle
\Bold{\mathrm{u}}\delta_{\Bold{\zeta}(\tau_{a'})},
\Bold{\zeta}\rangle\right\}}
\;d\Bold{\mathrm{u}}\,d\tau_{a'}\,\DO{\widetilde{\Omega}}\nonumber\\
&=&\int_{\Ta}\taurange\mathbf{1}
e^{\left\{\frac{-\pi|\x_{a'}-\x_a|^2}{s\tau_{a'}}\right\}}
\int_{\R^n}\exp{\left\{-s\pi
\widetilde{W}(\Bold{\mathrm{u}}\delta_{\Bold{\zeta}(\tau_{a'})})
\right\}}\,d\Bold{\mathrm{u}}\,d\tau_{a'}\DO{\tr}
\nonumber\\
&=&\int_{\Ta}\taurange\mathbf{1}e^{\left\{\frac{-\pi|\x_{a'}-\x_a|^2}
{s\tau_{a'}}\right\}}
\int_{\R^n}\exp^{\left\{-s\pi
\mathrm{u}^2\tau_{a'}\right\}}\,d\Bold{\mathrm{u}}\,d\tau_{a'}
\DO{\tr}\nonumber\\
&=&\int_{\Ta}\taurange\mathbf{1}e^{\left\{\frac{-\pi|\x_{a'}-\x_a|^2}
{s\tau_{a'}}\right\}}
(s\tau_{a'})^{-n/2}\,d\tau_{a'}\DO{\tr}\nonumber\\
&=&\taurange\mathbf{1}e^{\left\{\frac{-\pi|\x_{a'}-\x_a|^2}
{s\tau_{a'}}\right\}}
(s\tau_{a'})^{-n/2}\,d\tau_{a'}\;.
\end{eqnarray}
For $s=1$, this evaluates to
\begin{eqnarray}\label{infinite elliptic}
\K_\mathbb{U}^{(D)}(\x_a,\x_{a'})&=&\left\{\begin{array}{lr}
\pi^{1-n/2}\,\Gamma(\frac{n}{2}-1)|\x_{a'}-\x_a|^{2-n}\mathbf{1}
& \mbox{for $n\neq 2$}\\
-2\ln{|\x_{a'}-\x_a|}\mathbf{1} & \mbox{for $n=2$}
\end{array}\;,
\right.\nonumber\\
\end{eqnarray}
and for $s=i$,
\begin{equation}
\K_\mathbb{U}^{(D)}(\x_a,\x_{a'})=I(\x_a,\x_{a'};n)\mathbf{1}
\end{equation}
where
\begin{equation}
  I(\x_a,\x_{a'};n):=
  i^{-n/2}2^{n/2-1}\int_{\R}
  \mathrm{k}^{(n/2-2)}\exp{\left\{2\pi i\,\mathrm{k}|\x_{a'}-\x_a|^2
  \right\}}\,d\mathrm{k}
\end{equation}
after the change of integration variable
$(\tau_{a'})^{-1}\rightarrow 2\mathrm{k}$ with $\mathrm{k}$ real.
In particular, for $n=4$,
\begin{equation}
I(\x_a,\x_{a'};4)=2\delta[|\x_{a'}-\x_a|^2]\mathbf{1}\;.
\end{equation}

Note that the same answer obtains if the parametrization
$x(\tr,z)=\x_a+z(\tr)$ is used instead of the exponential map
parametrization. In this case, the action functional has no
$|\x_a-\x_{a'}|$ dependence, but the delta functional does.
Writing the delta functional as an integral and evaluating the
resulting functional integral yields the correct exponential
dependence on $|\x_a-\x_{a'}|$.

\subsubsection{Parabolic Case}\label{parabolic case}
Here the boundary segment at infinity is $\partial
\mathbb{U}^-$---not the cauchy surface. In this case, vanishing
Dirichlet boundary conditions necessarily implies
$\lim_{\x_{b}\rightarrow\infty}\Bold{\psi}(\x_{b},0)=0$ since
$\Bold{\psi}(\x_B,0)=\Bold{\phi}(\x_B,0)$ must hold for
consistency. The parametrization is the same as in the elliptic
case with the addition of $x^0(\tr)=\x_{b}^0-\tr$. The $x^0$
sector contributes nothing to $Q$ since $\int
|\dot{\Bold{x}^0}-\mathbf{Y}^0|^2 \,d\tau=0$. (This follows from
the parametrization (\ref{newparametrization}).)

The kernel associated with transitions from the Cauchy surface is
given by (\ref{second homogeneous K(D)}). In this example, it
satisfies
\begin{equation}
  \left.\left[\frac{s^2}{4\pi}\Delta-s\frac{\partial}{\partial
  \x^0}\right]\right|_{\x=(\x_a,\x_B^0)}
  \K_{C}^{(D)}((\x_{b},\x_{b}^0),(\x_a,0))=0\;.
\end{equation}
The calculation proceeds as before;
\begin{eqnarray}
  \K_{C}^{(D)}((\x_{b},\x_{b}^0),(\x_a,0))
  &=&\int_{T_{x_{cr}}\Z}
  \Bold{\delta}(\x_{b}\cdot\mathit{\Sigma}(\x_{b}^0,\Bold{\zeta})-\x_a)
 \exp\left\{-s^{-1}\pi \widetilde{Q}\right\}
  \;\mathcal{D}_{\widetilde{Q},\widetilde{W}}\Bold{\zeta}\nonumber\\
  &=&(s\x_{b}^0)^{-n/2}
  \exp{\left\{\frac{-\pi|\x_a-\x_{b}|^2}{s\x_{b}^0}\right\}}\mathbf{1}
\end{eqnarray}
for $\x_{b}^0\geq 0$.

For the elementary kernel find, using (\ref{interior K(D)}),
\begin{eqnarray}
  \K_\mathbb{U}^{(D)}((\x_{b},\x_{b}^0),(\x_a,\x_a^0))
  &=&\int_{\Ta}\int_{T_{x_{cr}}\Z}\taurange
  \Bold{\delta}(\x_{b}\cdot\mathit{\Sigma}(\tau_{a'},\Bold{\zeta})-\x_a)
  \delta(\x_{b}^0-\x_a^0-\tau_{a'})
  \nonumber \\
  &&\hspace{.6in}
  \times\exp\left\{-s^{-1}\pi \widetilde{Q}\right\}
  \;d\tau_{a'}\mathcal{D}_{\widetilde{Q},\widetilde{W}}\Bold{\zeta}
  \DO{\tr}
  \nonumber\\
  &=&\int_{\Ta}\taurange\mathbf{1}\delta(\x_{b}^0-\x_a^0-\tau_{a'})
  (s\tau_{a'})^{-n/2}
  e^{\left\{\frac{-\pi|\x_a-\x_{b}|^2}{s\tau_{a'}}\right\}}
  d\tau_{a'}\DO{\tr}\nonumber\\
  &=&\taurange\mathbf{1}\delta(\x_{b}^0-\x_a^0-\tau_{a'})
  (s\tau_{a'})^{-n/2}
  e^{\left\{\frac{-\pi|\x_a-\x_{b}|^2}{s\tau_{a'}}\right\}}
  \,d\tau_{a'}\nonumber\\
\end{eqnarray}
For $s=1$,
$\K_\mathbb{U}^{(D)}((\x_{b},\x_{b}^0),(\x_a,\x_a^0))=0$ for
$(\x_{b}^0-\x_a^0)<0$. If $(\x_{b}^0-\x_a^0)\geq 0$, then
\begin{eqnarray}
\K_\mathbb{U}^{(D)}((\x_{b},\x_{b}^0),(\x_a,\x_a^0))
  =(\x_{b}^0-\x_a^0)^{-n/2}
  \exp{\left\{\frac{-\pi|\x_a-\x_{b}|^2}{(\x_{b}^0-\x_a^0)}\right\}}
  \mathbf{1}\;.
\end{eqnarray}
For $s=i$,
\begin{equation}
  \K_\mathbb{U}^{(D)}((\x_{b},\x_{b}^0),(\x_a,\x_a^0))
  =[i(\x_{b}^0-\x_a^0)]^{-n/2}
  \exp{\left\{\frac{-\pi|\x_a-\x_{b}|^2}{i(\x_{b}^0-\x_a^0)}
  \right\}}
  \mathbf{1}
\end{equation}
for either $(\x_{b}^0-\x_a^0)\geq 0$ or $(\x_a^0-\x_{b}^0)\geq 0$.
Intuitively, for imaginary $s$, $\K_\mathbb{U}^{(D)}\rightarrow
{\K_\mathbb{U}^{(D)}}^*$ under time reversal.

\subsubsection{Hyperbolic Case}
Consider the wave equation. For this case $Q$ is imaginary with
index $(n-1,1)$. Choose $\mathbf{Y}=\mathbf{Y}^0=0$,
$\mathbf{X}^0=\partial/\partial \x_{b}^0$,
$\mathbf{X}_{\alpha}=\delta_{\alpha}^i\frac{\partial}{\partial
\x^i}$, and the coordinates in the obvious way so that the PDE for
the elementary kernel is
\begin{equation}\label{wave equation}
\frac{s^2}{4\pi}\left.\left[\Delta-\frac{\partial^2}{\partial
  {\x^0}^2}\right]\right|_{\x=(\x_{b},\x_{b}^0)}
  \K_{\mathbb{U}}^{(D)}((\x_{b},\x_{b}^0),(\x_a,\x_a^0))
  =-\Bold{\delta}((\x_{b},\x_{b}^0)-(\x_a,\x_a^0))
\end{equation}
where
\begin{eqnarray}\label{wave kernel}
  \K_\mathbb{U}^{(D)}((\x_{b},\x_{b}^0),(\x_a,\x_a^0))
  &=&\int_{\widetilde{\Omega}}\taurange
  \Bold{\delta}((\x_{b},\x_{b}^0)
  \cdot\mathit{\Sigma}(\tau_{a'},\Bold{\zeta})-(\x_a,\x_a^0))\nonumber \\
  &&\hspace{.6in}
  \times\exp\left\{-s^{-1}\pi \widetilde{Q}\right\}\;d\tau_{a'}
  \DO{\widetilde{\Omega}}\;.
\end{eqnarray}

The quadratic form is dictated by the form of the PDE to be
solved. It is more compact to express it in terms of coordinates
$\y\,^{\rho}=(\y^0,\y^1,\ldots,\y^{n-1})$ on
$\mathbb{E}^{(1,n-1)}$ with metric $\eta_{\rho\sigma}$:
\begin{eqnarray}
  \widetilde{Q}(y(\tau_{a'},\Bold{\zeta});s)&=&\int_0^{\tau_{a'}}
  \eta_{\rho\sigma}
  \dot{\Bold{y}}^{\rho}(\tau,z)\dot{\Bold{y}}^{\sigma}(\tau,z)d\tau
  \nonumber\\
  &=&\frac{(\y_a-\y_{a'})^2}{\tau_{a'}}+\int_0^{\tau_{a'}}
  \eta_{\alpha\beta}
  \dot{\Bold{\zeta}}^{\alpha}(\tau)\dot{\Bold{\zeta}}^{\beta}(\tau)d\tau\;.
\end{eqnarray}
The first term in the second equality comes from the critical path
in the same way as the elliptic case.

Repeating the calculation for the elliptic case yields
\begin{eqnarray}
  \K_\mathbb{U}^{(D)}(\y_{b},\y_a)
  &=&\int_{\Ta}\taurange\int_{\R^n}
  \mathbf{1}e^{\left\{\frac{\pi(\y_a-\y_{b})^2}{s\tau_{a'}}\right\}}
\exp^{\left\{-\pi
s\mathrm{u}^2\tau_{a'}\right\}}\,d\Bold{\mathrm{u}}
\,d\tau_{a'}\DO{\tr}\nonumber\\
&=&\int_{\Ta}\taurange\mathbf{1}
e^{\left\{\frac{\pi(\y_a-\y_{b})^2}{s\tau_{a'}}\right\}}
(s\tau_{a'})^{-n/2}\,d\tau_{a'}\DO{\tr}\nonumber\\
&=&\taurange\mathbf{1}
e^{\left\{\frac{\pi(\y_a-\y_{b})^2}{s\tau_{a'}}\right\}}
(s\tau_{a'})^{-n/2}\,d\tau_{a'}\nonumber\\
&=&\int_{\R}\mathbf{1} e^{\left\{\frac{-\pi i
(\y_a-\y_{b})^2}{\tau_{a'}}\right\}}
(i\tau_{a'})^{-n/2}\,d\tau_{a'}\nonumber\\
&=&I(\y_{b},\y_a;n)\mathbf{1}
\end{eqnarray}
where the second-to-last equality follows because
$\widetilde{Q}=-\widetilde{Q}^*$ allows $s=i$ only. In particular,
in $\E^{(1,3)}$,
\begin{eqnarray}
  \K_\mathbb{U}^{(D)}(\y_{b},\y_a)&=&2\delta[(\y_a-\y_{b})^2]\mathbf{1}\;.
\end{eqnarray}
 For
${\y^0}\,'>\y^0$,\footnote{Recall that either $\x_{b}^0\geq
\x_a^0$ or $\x_a^0\geq \x_{b}^0$.} this corresponds to the
retarded Green's function which has support on the future light
cone. Since the kernel is symmetric under $\y_a\leftrightarrow
\y_{a'}$, the same kernel obtains for $\y^0>{\y^0}\,'$ and
corresponds to the advanced Green's function.

\subsection{Planar Boundaries}\label{sec. planar boundaries}
For planar boundaries, the issue of dealing with $\taubot$ can be
side-stepped by employing a trick. Express $\mathbb{U}$ as
quotient space $\mathbb{U}=\widetilde{\mathbb{U}}/G$ where
$\widetilde{\mathbb{U}}$ is isomorphic to $\R^n$ and $G$ is the
discrete group required to obtain the desired boundary. It turns
out that the elementary kernels on $\mathbb{U}$ can be expressed
in terms of (presumably) known quantities on
$\widetilde{\mathbb{U}}$.

Let $\widetilde{\mathbb{U}}$ denote the universal covering of
$\mathbb{U}$. The manifold $\widetilde{\mathbb{U}}$ is a principal
fiber bundle with projection
$\Pi:\widetilde{\mathbb{U}}\rightarrow
\mathbb{U}=\widetilde{\mathbb{U}}/G$ and discrete structure group
$G$. The parametrization is constructed via the composition
$P=P_{\Pi}\circ\widetilde{P} $ where $\widetilde{P}:\Za\rightarrow
{\mathit{P}_a^{\widetilde{\mathbb{V}}}\widetilde{\mathbb{U}}}$ and
$P_{\Pi}:{\mathit{P}_a^{\widetilde{\mathbb{V}}}\widetilde{\mathbb{U}}}
\rightarrow{\mathit{P}_a^{\mathbb{V}}\mathbb{U}}$.

Given a connection on $\widetilde{\mathbb{U}}$ and a fixed initial
point $\x_a\in \mathbb{U}$, a given path $x(\cdot ,z)$ can be
horizontally lifted into the set of paths
$\{\widetilde{x}_{(p)}(\cdot ,z)\}$ with the index $p$ running
over the order of the discrete group. Hence, defining the
evaluation map $\varepsilon:\Za\rightarrow \mathbb{U}$ and its
lift $\widetilde{\varepsilon}:\Za\rightarrow
\widetilde{\mathbb{U}}$, then it follows that
$\varepsilon=\Pi\circ \widetilde{\varepsilon}$ and
$\widetilde{\varepsilon}^{-1}(\widetilde{x}_{(p)}(\cdot ,z))
\in{\Za}_{p}$ where ${\Za}_{p}$ parametrizes the $p\,\mathrm{th}$
path $\widetilde{x}_{(p)}(\cdot,z)$. Consequently, the inverse
image of the set of paths with a given base space end-point
consists of the disjoint union of inverse images of the set of
paths with end-points in the fiber over the base space end-point;
i.e., $\Za$ can be decomposed as $\bigcup_{p}{\Za}_{p}$ with
${\Za}_{p}\bigcap {\Za}_{q}=\varnothing\; \forall p\neq q$.

In Appendix \ref{app. universal covering}, it is shown that the
elementary kernel on $\mathbb{U}$ can be expressed in terms of the
elementary kernel on $\widetilde{\mathbb{U}}$ (provided it is an
equivariant map) by
\begin{equation}\label{covering space kernel}
    \K_\mathbb{U}^r(\x_a,\x_{a'})=
  \sum_p\rho^r(g_p^{-1})\widetilde{\K}_{\mathbb{U}(p)}
  (\x_a,\x_{a'})
\end{equation}
where the $\rho^r$ furnish unitary    representations---labelled
by $r$---of the structure group $G$, and
$\widetilde{\K}_{\mathbb{U}(p)}$ is the elementary kernel on
$\widetilde{\mathbb{U}}$ associated with the integral over
${\Za}_p$ and is given by
\begin{eqnarray}
\widetilde{\K}_{\mathbb{U}(p)}(\x_{a},\x_{a'})
&=&\int_{\Omega_{p}}\taurange
  \theta(\taubot-\tau_{a'})
  \widetilde{\Bold{\delta}}
(\x_{a}\cdot
[g_p^{-1}\widetilde{\mathit{\Sigma}}(\tau_{a'},z_p)g_p],
\x_{a'})\nonumber\\
  &&\hspace{.4in}\times
  \exp{\left\{-\mathcal{S}(x(\tau_{a'},z_p))\right\}}
  \;d\tau_{a'}\,\DO{\Omega_{p}}\nonumber\\
\end{eqnarray}
where $z_p\in{\Za}_p$ and $\Omega_p:={\Za}_p\times\Ta$.
Apparently, using this trick and variations of the following three
examples, one can find (at least in principle) the elementary
kernels of fairly general $n$-polygons (notwithstanding smoothness
issues on the boundary).

\subsubsection{The Half-Space}\label{sec. the half-space}
In this example, I treat only the parabolic case.  The elliptic
and hyperbolic cases can be obtained similarly. Define the
half-space
$\mathbb{U}^-=\R^{n}_+:=\{\x=(\x^1,\ldots,\x^n)\in\R^n|\;\x^n\geq
0\}$ so that $\widetilde{\mathbb{U}^-}=\R^n$. The boundary segment
$\partial \mathbb{U}^-$ is at infinity for points
$(\x^1,\ldots,\x^{n-1},0)$.

In this case, $G=\mathbb{Z}_2$ and the parametrization $P$ is
given by
\begin{equation}
  x(\tr,\Bold{\zeta})=(x^1_{cr}(\tr)+\zeta^1(\tr),\ldots,|x^n_{cr}(\tr)
  +\zeta^n(\tr)|)\;.
\end{equation}
Since $G=\mathbb{Z}_2$, the inverse image of a point $\x$ is
$\widetilde{\x}_{\pm}:=(\x^1,\ldots , \pm \widetilde{\x}\,^n)$
where the parametrization $\widetilde{P}$ gives
\begin{equation}
\pm \widetilde{x}\,^n(\cdot,z) =\pm (\x^n_{a'}+ z^n(\cdot))\;.
\end{equation}
Defining $\overline{\x}_{\pm}:=\{\x^1,\ldots,\pm \x^n\}$, the
delta functional contributes
\begin{equation}
  \widetilde{\Bold{\delta}}({\overline{\x}_{a'}}_{\pm}
  \pm z(\tau_{a})-\x_a)\;.
\end{equation}

Using the results of Subsection \ref{parabolic case} and
(\ref{covering space kernel}), the elementary kernel is
\begin{eqnarray}\label{half-space kernel}
\lefteqn{\K_\mathbb{U}^r((\x_{a'},\x_{a'}^0),(\x_a,\x_a^0))} \nonumber\\
&=& \rho^r(g_+)\widetilde{\K}_{\mathbb{U}(+)}
((\x_{a'},\x_{a'}^0),(\x_a,\x_a^0))
-\rho^r(g_-)\widetilde{\K}_{\mathbb{U}(-)}
((\x_{a'},\x_{a'}^0),(\x_a,\x_a^0))\nonumber\\
\end{eqnarray}
where
\begin{eqnarray}
\widetilde{\K}_{\mathbb{U}(\pm)}((\x_{a'},\x_{a'}^0),(\x_a,\x_a^0))
=[s(\x_{a'}^0-\x_a^0)]^{-n/2}
  \exp{\left\{\frac{-\pi| \x_{a}
  -{\overline{\x}_{a'}}_{\pm}|^2}{s(\x_{a'}^0-\x_a^0)}\right\}}
  \mathbf{1}\;.\nonumber\\
\end{eqnarray}
for real and imaginary $s$ (with the appropriate restrictions on
$(\x_{a'}^0-\x_a^0)$).

For the trivial representation,
$(\rho^{(D)}(g_+),\rho^{(D)}(g_-)):=(1,1)$, $\K_\mathbb{U}$
vanishes on the boundary $\x^n=0$. For the representation,
$(\rho^{(N)}(g_+),\rho^{(N)}(g_-)):=(1,-1)$,
$\nabla_{\mathbf{n}_{\partial}}\K_\mathbb{U}$ vanishes on the
boundary $\x^n=0$, and, therefore, these two representations yield
the Dirichlet and Neumann elementary kernels respectively.

Alternatively the Dirichlet and Neumann kernels for the half-space
can be obtained directly from Proposition 3.1 in \cite{LA2}. The
transformation $\sigma$ is just a translation of
$\x^n_{a'}\rightarrow 0$ so that $\x^n_{a}\rightarrow
\x^n_{a}+\x^n_{a'}$. Making use of Subsection \ref{sec. boundary
at infinity}, one easily arrives at (\ref{half-space kernel}).

The Cauchy kernel follows similarly;
\begin{eqnarray}
\lefteqn{\K_C^r((\x_{b},\x_{b}^0),(\x_a,0))} \nonumber\\
&=&\rho^r(g_+)\widetilde{\K}_{C(+)}((\x_{b},\x_{b}^0),(\x_a,0))
-\rho^r(g_-)\widetilde{\K}_{C(-)}((\x_{b},\x_{b}^0),(\x_a,0))
\end{eqnarray}
where
\begin{eqnarray}
\widetilde{\K}_{C(\pm)}((\x_{b},\x_{b}^0),(\x_a,0))
=(s\x_{b}^0)^{-n/2}
  \exp{\left\{\frac{-\pi| \x_{a}- {\overline{\x}_{b}}_{\pm}|^2}
  {s\x_{b}^0}\right\}}
  \mathbf{1}\;.
\end{eqnarray}
The Dirichlet and Neumann kernels obtain for the appropriate
representation of $\mathbb{Z}_2$.

To find the boundary kernel\footnote{Admittedly, the boundary
kernel can be obtained from the normal derivative of
$\K_{\mathbb{U}}^{(D)}$, but it is interesting to derive it from a
path integral.}, calculate
\begin{eqnarray}
 \K_{\partial{\mathbb{U}^-}}^{(D)}((\x_{b},\x_{b}^0),(\x_B,\x_B^0))
  &=&\int_{\Omega'}
  \Bold{\delta}((\x_{b}^i,\x_{b}^n)
  \cdot\mathit{\Sigma}(\taubotp,z),\x_B)
  \delta(\x_B^0-\x_{b}^0+\taubotp)
  \nonumber \\
  &&\hspace{.2in}
  \times\exp\left\{-s^{-1}\mathcal{S}((x,x^0)(\taubotp,z))\right\}
  \;\DO{\Omega'}
  \end{eqnarray}
where $\x_B=(\x_B^i,0)$ with $i\in\{1,\ldots,n-1\}$. The boundary
kernel vanishes for $\x_B^n\neq 0$. Since $\x_B^n=0$, the integral
over $\Z$ reduces to the unbounded case. Moreover, the $\Tb$
integral can be reduced according to (B.54) in \cite{LA2} since it
depends only on $\taubotp$;
\begin{eqnarray}
\K_{\partial{\mathbb{U}^-}}^{(D)}((\x_{b},\x_{b}^0),(\x_B,\x_B^0))
  &=&\int_{\Ta}\mathbf{1}
  (s\taubotp)^{-n/2}
  e^{\left\{\frac{-\pi|\x_B-\x_{b}|^2}{s\taubotp}\right\}}
  \delta(\x_B^0-\x_{b}^0+\taubotp)\;\DO{\tr}\nonumber\\
  &=&\mathcal{N}\int_{C_+}\mathbf{1}
  (s\taubotp)^{-(n/2+1)}
  e^{\left\{\frac{-\pi|\x_B-\x_{b}|^2}{s\taubotp}\right\}}
  \nonumber\\
  &&\hspace{1in}\times
  \delta(\x_B^0-\x_{b}^0+\taubotp)
  \;d\taubotp\nonumber\\
  &=&\mathcal{N}
  (s(\x_{b}^0-\x_B^0))^{-(n/2+1)}
  e^{\left\{\frac{-\pi|\x_B-\x_{b}|^2}{s(\x_{b}^0-\x_B^0)}\right\}}
  \mathbf{1}\nonumber\\
\end{eqnarray}
for $\x_{b}^0>\x_B^0$ and $\K_{\partial{\mathbb{U}^-}}^{(D)}=0$
otherwise.

To determine $\mathcal{N}$, normalize
$\K_{\partial{\mathbb{U}^-}}^{(D)}$ by choosing
$\Bold{\phi}(\x_B,\x_B^0)=a$ where $a$ is a constant. Then
\begin{eqnarray}
  \Bold{\Psi}^{(P)}(\x_{b},\x_{b}^0)
  &=&a\int_{\partial \mathbb{U}^-}\int_{I^-}
\K_{\partial{\mathbb{U}^-}}^{(D)}((\x_{b},\x_{b}^0),(\x_B,\x_B^0))
\;d\x_B\,d\x_B^0\nonumber\\
&=&a \mathcal{N}\int_0^{\x_{b}^0}
\mathbf{1}(s(\x_{b}^0-\x_B^0))^{-3/2}
e^{\left\{\frac{-\pi|\x_{b}^n|^2}{s(\x_{b}^0-\x_B^0)}\right\}}
  \;d\x_B^0\nonumber\\
  &=&a \mathcal{N}
  (s\x_{b}^n)^{-1}\left[\mathrm{erf}
  \left\{\sqrt{\frac{\pi}{s\x_{b}^0}}|\x_{b}^n|\right\}-1\right]
  \mathbf{1}\;.
\end{eqnarray}
Since $\Bold{\Psi}^{(P)}(\x_B,\x_{b}^0)=a$ and
$\mathrm{erf}(0)=0$, then $\mathcal{N}=-s\x_{b}^n$. Finally,
  \begin{eqnarray}
\K_{\partial{\mathbb{U}^-}}^{(D)}((\x_{b},\x_{b}^0),(\x_B,\x_B^0))
    =-s\x_{b}^n
  (s(\x_{b}^0-\x_B^0))^{-(n/2+1)}
  e^{\left\{\frac{-\pi|\x_B-\x_{b}|^2}{s(\x_{b}^0-\x_B^0)}\right\}}
  \mathbf{1}\nonumber\\
\end{eqnarray}
for $\x_{b}^0>\x_B^0$ and $\K_{\partial{\mathbb{U}^-}}^{(D)}=0$
otherwise.

More generally, this procedure can be applied to $\R^n$ ``folded''
into $1/(2m)$-space ($m\leq n$) by the structure group
$\underbrace{\mathbb{Z}_2\oplus,\ldots ,\oplus\mathbb{Z}_2}_m$
where each ``fold'' is along a different $\x^i$. Another
possibility is to use $\mathbb{Z}_h$ with $h$ a positive integer
with $\mathbb{U}=\R^2$ to obtain a sector in the plane.

\subsubsection{The Unit Strip}
Consider only the elliptic elementary kernel case for brevity.
Define the unit strip by
$\mathbb{U}\equiv\{\x=(\x^1,\ldots,\x^n)\in\R^n|\;0\leq \x^n\leq
1$. An appropriate parametrization given the initial point
$\x_a=(\x_a^1,\ldots, |\sin(\mathrm{k}\x_a^n)|)=:(\x_a^1,\ldots,
\widehat{\x}^n_a)$ is
\begin{equation}
  x(\tr,\Bold{\zeta})
  =(x^1_{cr}(\tr)+\zeta^1(\tr),\ldots,|\sin{[\mathrm{k}(x^n_{cr}(\tr)
  +\zeta^n(\tr))]}|)
\end{equation}
where $x_{cr}(\tr)$ is given by (\ref{critical path}) and
$\mathrm{k}$ is a real constant.  The structure group is
$G=\mathbb{Z}_2\oplus\mathbb{Z}$ since the set of inverse images
is $\{\widetilde{\x}_{(\pm,m)}\}=(\x^1,\ldots
,\pm\widetilde{\x}_m^n)$ where
\begin{equation}
  \pm\widetilde{x}^n_m(\cdot,z)=\pm(\widehat{\x}_a^n+z^n(\cdot)\pm 2\pi
  m/\mathrm{k})
  \end{equation}
with $m/\mathrm{k}\in\mathbb{Z}$. The delta functional is
\begin{equation}
 \widetilde{\Bold{\delta}}({\overline{\x}_{a}}_{(\pm,m)}\pm z(\tau_{a'})
 -\x_{a'})
\end{equation}
with ${\overline{\x}}_{(\pm,m)}:=(\x^1,\ldots,\pm
(\widehat{\x}^n\pm 2\pi m/\mathrm{k}))$.

Repeating the relevant steps from Subsection \ref{elliptic case},
find the elementary kernel;
\begin{eqnarray}
\lefteqn{\widetilde{\K}_{\mathbb{U}(\pm,m)}(\x_a,\x_{a'})
  }\nonumber\\\nonumber\\
  &=&\left\{\begin{array}{ll}
\pi^{1-n/2}\,\Gamma(\frac{n}{2}-1)
|\x_{a'}-{\overline{\x}_{a}}_{(\pm,m)}|^{2-n}\mathbf{1}
& \mbox{for $n\neq 2$}\\
-2\ln{|\x_{a'}-{\overline{\x}_{a}}_{(\pm,m)}|}\mathbf{1} &
\mbox{for $n=2$}
\end{array}
\right.,\nonumber\\\nonumber\\
&=&I({\overline{\x}_{a}}_{(\pm,m)},\x_{a'};n)\mathbf{1}
\end{eqnarray}
for $s=1$ and $s=i$ respectively. Finally, for the trivial
representation of $\mathbb{Z}$, the Dirichlet and Neumann
elementary kernels are
\begin{eqnarray}
\K_\mathbb{U}^{\mp}(\x_a,\x_{a'})
=\sum_m\left[\widetilde{\K}_{\mathbb{U}(+,m)}(\x_a,\x_{a'})\mp
\widetilde{\K}_{\mathbb{U}(-,m)}(\x_a,\x_{a'})\right]\;.\nonumber\\
\end{eqnarray}

\subsubsection{The Unit Box}
Let $\mathbb{U}$ be the box defined by
$\mathbb{U}\equiv\{\x=(\x^1,\ldots,\x^n)\in\R^n|\;0\leq \x^i\leq
1\;\forall i\in\{1,\ldots ,n\}\}$. I will calculate only the
elementary kernel for the elliptic case again for brevity. For the
initial point $\widehat{\x}_a:=(|\sin(\mathrm{k}_1\x^1_a)|,\ldots
,|\sin(\mathrm{k}_n\x^n_a)|$, the parametrization is
\begin{equation}
  x(\tr,\Bold{\zeta})=(|\sin{[\mathrm{k}_1(x^1_{cr}(\tr)+\zeta^1(\tr))]}|,
  \ldots ,
  |\sin{[\mathrm{k}_n(x^n_{cr}(\tr)+\zeta^n(\tr))]}|)
\end{equation}
with $x_{cr}(\tr)$ given by (\ref{critical path}). The structure
group is
\begin{equation}
  G=\underbrace{\mathbb{Z}_2\oplus\mathbb{Z},\oplus,\ldots ,
  \oplus\mathbb{Z}_2\oplus\mathbb{Z}}_n\;,
\end{equation}
and the set of inverse images is
\begin{equation}
  \{\widetilde{\x}_{(\pm,m)_j}\}=(\pm \widetilde{\x}^1_{m_1},\ldots
,\pm \widetilde{\x}_{m_n}^n)
\end{equation}
where
\begin{equation}
  \pm\widetilde{x}^i_{m_i}(\cdot,z)
  =\pm(\widehat{\x}_a^i+z^i(\cdot)\pm 2\pi m_i/\mathrm{k}_i)
\end{equation}
with $i\in \{1,\ldots ,n\}$ and $m_j/\mathrm{k}_j\in \mathbb{Z}$.

Proceeding as in the previous examples, find the kernel for $s=1$
and $s=i$ respectively:
\begin{eqnarray}
\lefteqn{\widetilde{\K}_{\mathbb{U}(\pm,m)_j}(\widehat{\x}_a,\x_{a'})
  }\nonumber\\\nonumber\\
  &=&\left\{\begin{array}{ll}
\pi^{1-n/2}\,\Gamma(\frac{n}{2}-1)
|\x_{a'}-{\overline{\x}_a}_{(\pm,m)_j}|^{2-n}\mathbf{1}
& \mbox{for $n\neq 2$}\\
-2\ln{|\x_{a'}-{\overline{\x}_a}_{(\pm,m)_j}|}\mathbf{1} &
\mbox{for $n=2$}
\end{array}
\right.,\nonumber\\\nonumber\\
&=&I({\overline{\x}_a}_{(\pm,m)_j},\x_{a'};n)\mathbf{1}
\end{eqnarray}
where ${\overline{\x}}_{(\pm,m)_j}:=\{\pm(\widehat{\x}\,^1\pm 2\pi
m_1/\mathrm{k}_1),\ldots,\pm(\widehat{\x}\,^n\pm 2\pi
m_n/\mathrm{k}_n)\}$. This yields the Dirichlet and Neumann
elementary kernels (for trivial representations of the
$\mathbb{Z}$)
\begin{eqnarray}
\K_\mathbb{U}^{\mp}(\widehat{\x}_a,\x_{a'})
=\sum_{\{m_{j}\}}\left[\widetilde{\K}_{\mathbb{U}(+,m)_j}
(\widehat{\x}_a,\x_{a'})\mp
\widetilde{\K}_{\mathbb{U}(-,m)_j}(\widehat{\x}_a,\x_{a'})\right]
\end{eqnarray}
where $\{m_{j}\}:=\{m_1,\ldots,m_n\}$.

\subsection{Spherical Boundaries in $\R^n$}\label{sec. spherical
boundaries}
\subsubsection{The $n$-Ball}\label{sec. the n-ball}
To handle spherical boundaries, it is necessary to come to grips
with $\taubot$. The calculations for the boundary kernels can be
simplified substantially by expanding the paths about a critical
path and then using (3.33) in \cite{LA2} which allows the point
$x_a^\bot$ to be replaced by an arbitrary $x_B$ in the boundary
kernel.

Consider Laplace's equation for the case of the $n$-ball in $\R^n$
defined by
$\mathbb{B}^n:=\{\x=(\x^1,\ldots,\x^n)\in\R^n|\,\x^2-R^2\leq 0\}$.
First calculate $\K_{\partial}^{(D)}(\x_a,\x_a^\bot)$. As usual,
it is expedient to expand about a critical path and the
exponential map parametrization is
\begin{equation}
  x^2(\tr,\Bold{\zeta})=\left[x_{cr}(\tr)+\Bold{\zeta}(\tr)\right]^2
\end{equation}
where
\begin{equation}
  x_{cr}(\tr)=\x_a\left(1-\frac{\tr}{\taubot}\right)
  +\frac{\x_a^\bot}{\taubot}\,\tr
\end{equation}
is a critical path.

The form of $x_{cr}$ is dictated by the initial condition of the
parametrization, which applies to $x(\tr,\Bold{\zeta})$; and by
the transversality condition (see Section 2 in \cite{LA2}), which
requires that $\dot{\Bold{x}}_{cr}(\taubot)$ intersects the
boundary transversally. A reasonable choice is
$|\dot{\Bold{x}}_{cr}(\taubot)|\propto|\x_a^\bot|/\taubot$. Note,
however, that the transversality condition does not fix the
proportionality constant. The proportionality constant is
ultimately fixed by requiring that
$\K_{\partial}^{(D)}(\x_{B'},\x_a^\bot)=\Bold{\delta}(\x_{B'}-\x_a^\bot)$.

Now evaluate
\begin{equation}
  \K_{\partial}^{(D)}(\x_a,\x_a^\bot)=\int_{\widetilde{\Omega}}
  \Bold{\delta}(x(\taubot,\Bold{\zeta})-\x_a^\bot)
  \exp{\left\{-s^{-1}\pi \widetilde{Q}\right\}}\;\DO{\widetilde{\Omega}}\;,
\end{equation}
where the quadratic form is
\begin{eqnarray}
  \widetilde{Q}(\Bold{\zeta};\taubot)&=&\int_0^{\taubot}
  \delta_{ij}\dot{\Bold{x}}^{i}(\tau,z)\dot{\Bold{x}}^{j}(\tau,z)d\tau
  \nonumber\\
  &=&\frac{|\x_a^\bot-\x_a|^2}{\taubot}+\int_0^{\taubot}
  \delta_{\alpha\beta}
  \dot{\Bold{\zeta}}^{\alpha}(\tau)\dot{\Bold{\zeta}}^{\beta}(\tau)d\tau\;.
\end{eqnarray}
The first step is to note that the integrand depends on $\taubot$
only. Therefore, it reduces to an integral over $C_+$ as usual.
Using (B.54) in \cite{LA2}, get
\begin{equation}
  \K_{\partial}^{(D)}(\x_a,\x_a^\bot)
  = \mathcal{N}\int_{C_+}\int_{\Za}
  \Bold{\delta}(x(\taubot,\Bold{\zeta})-\x_a^\bot)
  \exp{\left\{-s^{-1}\pi \widetilde{Q}\right\}}\;\DQ{\Bold{\zeta}}\frac{d\taubot}{\taubot}\;.
\end{equation}
The integral over $\Za$ can be done as before since, by the
parametrization, $\Bold{\zeta}(\taubot)=0$. Use (3.34) in
\cite{LA2} to replace the point $\x_a^\bot$ with $\x_B$ to get
 \begin{eqnarray}\label{ball kernel}
 \K_{\partial}^{(D)}(\x_a,\x_B)
  &=&\int_{C_+}\mathbf{1}\frac{\mathcal{N}}{(\taubot)^{(n+2)/2}}\,
  \exp{\left\{-\pi
  \frac{|\x_B-\x_a|^2}{\taubot}\right\}}\;d\taubot
  \nonumber\\
  \nonumber\\
  &=&\left\{
  \begin{array}{ll}
  \mathcal{N}\frac{\Gamma(\frac{n}{2})}{\pi^{n/2}}|\x_B-\x_a|^{-n}\mathbf{1}
  \hspace{.8in}\mbox{for $s=1$}\\
  \mathcal{N}I(\x_a,\x_B;n+2)\mathbf{1}\,\hspace{.8in}\mbox{for $s=i$}
  \end{array}
  \right.\;.\nonumber\\
\end{eqnarray}
Find $\mathcal{N}$ by normalizing $\K_{\partial}^{(D)}$. For
$s=1$, integrating over $\partial \mathbb{B}^n$ and setting the
result equal to unity gives
\begin{equation}
\mathcal{N}=\frac{R^2-\x_a^2}{2R} \;.
\end{equation}
Consequently obtain the well known result,\footnote{Incidently,
this example shows that, for the $n$-ball,
$\langle\tau_{\mathrm{x}_a}^\bot\rangle_{\Ta}=(R^2-\x_a^2)/2$\;.}
\begin{equation}
  \K_{\partial}^{(D)}(\x_a,\x_B)
  =\frac{\Gamma(\frac{n}{2})}{2\pi^{n/2}}\,\frac{(R^2-\x_a^2)}
  {R|R\mathbf{n}_{\x_B}-\x_a|^n}
  \mathbf{1}\;.
\end{equation}

Now, for $\Bold{K}_{\mathbb{U}}^{(D)}$ the parametrization for
interior points becomes
\begin{equation}
  x^2(\tr,z)=\left[x_{cr}(\tr)+\Bold{\zeta}(\tr)\right]^2
\end{equation}
where $x_{cr}$ is given by (\ref{critical path}). Do the $\Za$
integral first to get,
\begin{eqnarray}\label{n-ball}
\K_\mathbb{U}^{(D)}(\x_a,\x_{a'})
&=&\int_{\Ta}\int_0^{\taubot}\mathbf{1}(s\tau_{a'})^{-n/2}
e^{\left\{\frac{-\pi|\x_{a'}-\x_a|^2}{s\tau_{a'}}\right\}}
\,d\tau_{a'}\,\DO{\tr}\;.
\end{eqnarray}
Now restrict to the $s=1$, $n=3$ case and utilize the
decomposition from Subsection 3.2.1 in \cite{LA2}.\footnote{For
comparison purposes, this is more convenient than solving
(\ref{n-ball}) directly (which entails finding a suitable
normalization constant $\mathcal{N}$ after localizing the integral
over $\Ta$).} In spherical coordinates, the transformation
$\sigma:r_a\rightarrow \alpha r_a$ where $\alpha=R/r_a$ takes
$\taubot\rightarrow 0$.\footnote{The transformation $\sigma$ must
be multiplicative here and not additive as it was for the planar
case: Additive would lead to $r_{a'}\leq 0$ in general, but
$r_{a'}>0$ by assumption.} Moreover, there is only one critical
path since $r_a\neq 0$ excludes the path that passes throught the
origin and intersects the boundary transversally. This yields
\begin{eqnarray}
 \mathbf{F}_\mathbb{U}(R\,\mathbf{n}_{\x_a},(|\x_a|/R)\x_{a'})
 &=&\int_{0}^{\infty}
  \mathbf{1}(\tau_{a'})^{-3/2}e^{\left\{\frac{-\pi|(|\x_a|/R)\x_{a'}
  -R\,\mathbf{n}_{\x_a}|^2}{\tau_{a'}}\right\}}
\,d\tau_{a'}\nonumber\\
&=&\frac{1}{\left|\frac{|\x_a|\x_{a'}}{R}-
R\,\mathbf{n}_{\x_a}\right|}\mathbf{1}\;.
\end{eqnarray}
The final result is
\begin{equation}
\K_\mathbb{U}^{(D)}(\x_a,\x_{a'})=
\left[\frac{1}{|\x_{a'}-\x_a|}-\frac{1}{\left|\frac{|\x_a|\x_{a'}}{R}-
R\,\mathbf{n}_{\x_a}\right|}\right]\mathbf{1}\;.
\end{equation}
It is interesting to compare and contrast this strategy to the
method of images.

For the exterior of the $n$-ball, the calculation for the
elementary kernel goes through as before. However, the boundary
kernel changes sign because the normalization factor is found to
be $\mathcal{N}=(\x_a^2-R^2)/2R$ since $|\x_a|>R$ in this case.

\subsubsection{Topological $n$-Ball}

The procedure employed in the previous subsection can be used to
determine the elementary kernel for more general boundaries  in
$\R^n$ that are topologically equivalent to $\mathbb{B}^n$. Quite
generally\footnote{I only exhibit the expressions for the case
$s=1$, but the general case can be handled analogously.},
\begin{eqnarray}
\K_\mathbb{U}^{(D)}(\x_a,\x_{a'})
&=&\int_{\Ta}\int_0^{\taubot}\mathbf{1}(\tau_{a'})^{-n/2}
e^{\left\{\frac{-\pi|\x_{a'}-\x_a|^2}{\tau_{a'}}\right\}}
\,d\tau_{a'}\DO{\tr}\nonumber\\
&=&\int_0^{\infty}\mathbf{1}(\tau_{a'})^{-n/2}
e^{\left\{\frac{-\pi|\x_{a'}-\x_a|^2}{\tau_{a'}}\right\}}
\,d\tau_{a'}\nonumber\\
&&-\int_{\Ta}\int_{\taubot}^{\infty}\mathbf{1}(\tau_{a'})^{-n/2}
e^{\left\{\frac{-\pi|\x_{a'}-\x_a|^2}{\tau_{a'}}\right\}}
\,d\tau_{a'}\DO{\tr}\;.
\end{eqnarray}
One recognizes the second term as a homogeneous term which ensures
that the kernel $\K_\mathbb{U}^{(D)}(\x_a,\x_{a'})$ has the
required boundary condition.

As in the previous subsection, make a
transformation(s)\footnote{Recall that one should sum over all
$\tau_{\mathrm{x}_a}^\bot$ if there are multiple critical paths.
Consequently there may be multiple tansformations as well.}
$\sigma:\x_a\rightarrow \x_a^\bot$. Then
$\tau_{\mathrm{x}_a}^\bot\rightarrow 0$ and end up with an
expression that is independent of $\tau_{\mathrm{x}_a}^\bot$;
\begin{eqnarray}
\K_\mathbb{U}^{(D)}(\x_a,\x_{a'})
&=&\int_0^{\infty}\mathbf{1}(\tau_{a'})^{-n/2}
\exp{\left\{\frac{-\pi|\x_{a'}-\x_a|^2}{\tau_{a'}}\right\}}
\,d\tau_{a'}\nonumber\\
&&-\int_{0}^{\infty} \mathbf{1}(\tau_{a'})^{-n/2}
\exp{\left\{\frac{-\pi|\sigma^{-1}(\x_{a'})
  -\sigma(\x_a)|^2}{\tau_{a'}}\right\}}
\,d\tau_{a'}\;.
\end{eqnarray}
Of course, the transformation(s) $\sigma$ can be quite complicated
in general and numerical methods may be required, but the solution
can be obtained in principle.

\subsection{The Quadrant}
As a final example, consider the upper right quadrant in $\R^2$,
and calculate $\K_\mathbb{U}^{(D)}(\x_a,\x_{a'})$ and
$\K_\p^{(D)}(\x_a,\x_{a'})$ for the elliptic case with $s=1$. This
problem can be handled by the technique of Subsection \ref{sec.
planar boundaries}, but it is instructive to solve it directly
because it illustrates the case when there are multiple critical
paths.

Denote a point $\x\in\R^2$ by $\x=(\x^1,\x^2)$. The elementary
kernel is given by
\begin{eqnarray}\label{quadrant kernel}
\K_\mathbb{U}^{(D)}(\x_a,\x_{a'})&=&\int_\Omega
\left[\int_0^{\infty} \Bold{\delta}(x(\tau_{a'},z),\x_{a'})
\exp{\left\{-\mathcal{S}(x(\tau_{a'},z))\right\}}\right.d\tau_{a'}
\nonumber\\
  &&\hspace{.2in}-\left.\int_{\taubot}^{\infty}
\Bold{\delta}(x(\tau_{a'},z),\x_{a'})
\exp{\left\{-\mathcal{S}(x(\tau_{a'},z))\right\}}d\tau_{a'}\right]
\;\DO{\Omega}\;.\nonumber\\
\end{eqnarray}
The first integral was already calculated in (\ref{infinite
elliptic});
$-2\ln|((\x_{a'}^1-\x_a^1),(\x_{a'}^2-\x_a^2))|\mathbf{1}$. To
calculate the second integral, note that there are three critical
paths: a perpendicular line from $\x_a$ to the $\x^1$ axis; a
perpendicular line from $\x_a$ to the $\x^2$ axis; and, because
the tangent along the boundary vanishes at the origin, a straight
line from $\x_a$ to the origin.

The transformations that take $\x_a$ to each $\x_a^\bot$ are:
$\sigma_1:(x^1_a,x^2_a)\rightarrow (0,x^2_a)$;
$\sigma_2:(x^1_a,x^2_a)\rightarrow (x^1_a,0)$; and
$\sigma_3:(x^1_a,x^2_a)\rightarrow (0,0)$. The hessian of
$\mathcal{S}(x_{cr})$ is degenerate at the points $(0,\x_{a'}^2)$,
$(\x_{a'}^1,0)$, and $(0,0)$. Consequently, care must be exercised
when evaluating
$\mathbf{F}_\mathbb{U}(\sigma(\x_a),\sigma^{-1}(\x_{a'}))$. In
fact the hessian vanishes at $(0,0)$, which is therefore a point
of nullity $2$, and one might guess that the critical path
intersecting the origin will pick up a phase $\exp(i\pi)$.

Instead of going into details, it suffices to indicate why this
expectation is reasonable and then simply check that it gives the
answer with the correct boundary conditions. The integral to be
evaluated is
\begin{eqnarray}
  \mathbf{F}_\mathbb{U}(0,\x_{a'}+\x_a)&=&
  \int_{0}^{\infty}\mathbf{1}(\tau_{a'})^{-1}
e^{\left\{\frac{-\pi|\x_{a'}+\x_a|^2}{\tau_{a'}}\right\}}
\,d\tau_{a'}\nonumber\\
&=&\int_{0}^{\infty}\mathbf{1}(\tau_{a'})^{-1}
e^{\left\{\frac{-\pi|\x_{a'}-(-\x_a)|^2}{\tau_{a'}}\right\}}
\,d\tau_{a'}\;.
\end{eqnarray}
Considered as an integral over the complex $\tau_{a'}$ plane, this
is just $\K_\mathbb{U}(\x_a,\x_{a'})$ with $\x_a$ rotated through
$\theta=\pm\pi$; indicating that $\mathbf{F}_\mathbb{U}$ should
pick up a phase of $\exp(\mp i\pi)$ relative to $\K_\mathbb{U}$
due to the $(\tau_{a'})^{-1}$ factor in the integrand.

Collecting the four contributions to (\ref{quadrant kernel})
yields
\begin{eqnarray}
\K_\mathbb{U}^{(D)}(\x_a,\x_{a'})
&=&-2\left\{\ln|((\x_{a'}^1-\x_a^1),(\x_{a'}^2-\x_a^2))|
-\ln|((\x_{a'}^1+\x_a^1),(\x_{a'}^2-\x_a^2))|\right.\nonumber\\
&&-\left.\ln|((\x_{a'}^1-\x_a^1),(\x_{a'}^2+\x_a^2))|
+\ln|((\x_{a'}^1+\x_a^1),(\x_{a'}^2+\x_a^2))|\right\}\mathbf{1}\;.
\nonumber\\
\end{eqnarray}
The first term satisfies the inhomogeneous equation, the next
three terms verify the homogeneous equation, and
$\left.\K_\mathbb{U}^{(D)}(\x_a,\x_{a'})\right|_{\x_a^1=0}
=\left.\K_\mathbb{U}^{(D)}(\x_a,\x_{a'})\right|_{\x_a^2=0}=0$.

Now, for the boundary kernel there are three $\taubot$
corresponding to the three critical paths. For each critical time,
it is best to expand about the associated critical path. For
example, expand about
\begin{equation}
  \left\{\begin{array}{ll}
  x_{cr}^1(\tau)=\x_a^1(1-\tau/\taubot)\\
  x_{cr}^2(\tau)=\x_a^2(1-\tau/\taubot)+x_{a'}^2(\tau/\taubot)
  \end{array}\right.
\end{equation}
for the critical time from $\x_a$ to the $x^1$ axis. Each critical
path will contribute to a segment of the boundary. Hence, the path
integral separates into three terms. The integral has already been
evaluated in (\ref{ball kernel}) (for a different critical path).
Use the three parametrizations for the critical paths to get
\begin{eqnarray}
\K_\p^{(D)}(\x_a,(\x_{a'}^1,0))&=&\mathcal{N}_1\pi^{-1}
  [(x_{a'}^1-x_a^1)^2+(x_a^2)^2]^{-1}\nonumber\\
  \K_\p^{(D)}(\x_a,(0,\x_{a'}^2))&=&\mathcal{N}_2\pi^{-1}
  [(x_a^1)^2+(x_{a'}^2-x_a^2)^2]^{-1}\nonumber\\
  \K_\p^{(D)}(\x_a,(0,0))&=&\mathcal{N}_3\pi^{-1}
  [(x_a^1)^2+(x_a^2)^2]^{-1}\;.
\end{eqnarray}
Finally, integrate each over its relevant range to normalize
$\K_\p^{(D)}$;
\begin{equation}
  \mathcal{N}_1=\frac{\x_a^2}{2[\tan^{-1}(\x_a^1/\x_a^2)+\pi/2]}\;;
  \hspace{.2in}
  \mathcal{N}_2=\frac{\x_a^1}{2[\tan^{-1}(\x_a^2/\x_a^1)+\pi/2]}\;;
  \hspace{.2in}\mathcal{N}_3=0\;.
\end{equation}

Similar reasoning can be applied to the wedge problem where the
angle is no longer restricted to $\pi/2$.

\section{Conclusion}
Specialization of the general path integral constructed in
\cite{LA2} leads to path integral solutions of elliptic,
parabolic, and hyperbolic PDEs with Dirichlet/Neumann boundary
conditions. The path integral, along with some calculational
techniques, was used to evaluate selected kernels of some known
planar and spherical boundary problems in $\R^n$. The techniques
for calculating the kernels for the $n$-ball and quotient spaces
are new to my knowledge, and they seem to be useful for evaluating
kernels with more complicated geometries. It may be useful to try
to solve for the kernels without recourse to the techniques
developed here by applying a perturbation technique to $\taubot$.
Non-vanishing $V(\x)$ can also be included and the resulting
integrals can be solved exactly in select cases or by expansion
techniques. It would be useful to do more substantial calculations
of kernels on non-trivial manifolds. This would require
parametrizing the paths using the Cartan development map and,
probably, new calculational techniques.

\vspace{.3in} \noindent\textbf{Acknowledgment} I thank C.
DeWitt-Morette for helpful suggestions and discussions.

\appendix
\section{Exponential Map Parametrization}\label{app. exp}
There are two maps that are typically used to parametrize paths on
a manifold; the Cartan development map and the exponential map.
The Cartan development map is presented in detail in
\cite{D-W/MA/NE}. The exponential map is particularly useful for
parametrizing paths on linear spaces and compact Lie groups.

The exponential map, $\mathit{Exp}:T_{x_f}\PM\rightarrow\PM$, is a
frequently used method of parametrization that yields the usual
loop expansion when $T\M$ is a linear space. It maps the tangent
space of the space of paths $\PM$ at some fiducial path $x_f$ to
the space of $L^{2,1}$ paths on $\M$ with fixed end-point at
$\ti_a$.

Assume the tangent space $T_{x_f}\PM$ is generated from paths
$x_f(\lambda,\ti)$ representing one-parameter variations of paths
in $\M$ with fixed end-point at $\ti_a$. Each variation yields a
vector field $\Bold{\zeta}=dx_f(\lambda)/d\lambda|_{\lambda =0}$
along $x_f(\ti)$. The vector $\Bold{\zeta}_{x_f}\in T_{x_f}\PM$ is
a map $\Bold{\zeta}_{x_f}:\mathbb{T}\rightarrow T_{x_f(\ti)}\M$
such that $\Bold{\zeta}_{x_f}(\ti_a)=0$.

The exponential map $\mathit{Exp}$ is defined point-wise on the
tangent bundle in terms of the map
$\mathrm{exp}:T_{x(t)}\M\rightarrow \M$. For each
$\Bold{\zeta}_{x_f}(\ti)\in T_{x_f(\ti)}\M$, the corresponding
family of paths $x(\ti,\Bold{\zeta};\lambda)\in \M$ is determined
by the differential equation
\begin{equation}
  \left.\frac{dx(\ti;\lambda)}{d\lambda}\right|_{\lambda=0}
  =\Bold{\zeta}_{x_f}(x(\ti))\hspace{.2in}x(\ti;0)=x_f(\ti)
\end{equation}
with the additional requirement
$x(\ti_a;\lambda)=x_f(\ti_a)\;\forall \lambda\in [0,1]$. The
solution in an open set $\mathbb{U}\subseteq \M$ can be formally
written as
\begin{equation}
  x(\ti,\Bold{\zeta};\lambda)=
  \exp\left(\lambda\Bold{\zeta}_{x_f}(\ti)\right)x_f(\ti)\;.
\end{equation}
Of course, the global uniqueness and existence of the solution
depends on the nature of $T\mathbb{U}\subseteq T\M$. This is a
particularly useful parametrization when $\M$ is a compact Lie
group.

If $T\mathbb{U}$ is a linear space, then the parametrization
reduces to
\begin{equation}
  x(\ti,\Bold{\zeta};\lambda)=x_f(\ti)+
 \lambda\Bold{\zeta}_{x_f}(\ti)\;.
\end{equation}
This is the parametrization used in the examples of Section
\ref{sec. examples} with $\lambda=1$ and $x_f=x_{cr}$ since
$T\mathbb{U}\equiv\R^{2n}$. If $\lambda$ is a small parameter,
then one replaces $x(\ti,\Bold{\zeta};\lambda)$ with $x_f(\ti)+
 \lambda\Bold{\zeta}_{x_f}(\ti)$ in the path integral and expands in
 powers of $\lambda$ which leads to a loop expansion in the path integral.

\section{Quotient Spaces}\label{app. universal covering}
When the manifold $\M$ can be contructed as the base space of a
principal fiber bundle with a discrete structure group, the
kernels on $\M$ can be expressed in terms of the kernels on the
fiber bundle. The first realization of this fact came in
\cite{LA/D-M} with a treatment of multiply connected spaces. Their
result can be trivially extended to quotient spaces with a
principal bundle structure. The construction is sketched in
\cite{DO} and \cite{CA/D-M}: The details are included here for
completeness. For a system with symmetry, the result allows the
kernels on the base space to be expressed in terms of kernels on
the principle bundle. In particular, using this result, it is
possible to express kernels on multiply connected spaces,
orbifolds, compact Lie groups, and homogeneous spaces in terms of
kernels on associated covering spaces.

Let the manifold $\widetilde{\M}$ be a principal fiber bundle
endowed with a connection, a projection
$\Pi:\widetilde{\M}\rightarrow \M$, and a structure group $G$. (If
$G\cong \pi_1(\M)$ this yields the multiply connected case.) Given
a set of open coverings $\{\mathbb{U}_i\}$ choose a trivialization
$\{\mathbb{U}_i,\overline{\varphi}_i\}$ of $\widetilde{\M}$ and
the canonical section
$\overline{s}_i=\overline{\varphi}_i^{-1}\circ\overline{Id}$ where
$\overline{Id}:\mathbb{U}_i\rightarrow \mathbb{U}_i\times G$ and
$\overline{\varphi}_i:\Pi^{-1}(\mathbb{U}_i)\rightarrow
\mathbb{U}_i\times G$. Then, under the trivialization,
$\widetilde{\x}=\overline{s}(\x)\cdot g\mapsto (\x,g)$ for
$\widetilde{\x}\in \widetilde{\M}$, $\x\in \M$, and $g\in G$.

Now let $T_{(p,q)}\M$ denote the tensor bundle weakly associated
to $\widetilde{\M}$. With a choice of trivialization
$s_i=\varphi_i^{-1}\circ Id$ on $T_{(p,q)}\M$, define an
equivariant $r$-form $\Bold{\widetilde{\Psi}}
:\widetilde{\M}\rightarrow(\mathbb{C}^m)^{p+q}$ by
\begin{equation}\label{important relation}
  \Bold{\widetilde{\Psi}}(\widetilde{\x}g)
  =\rho(g^{-1})\Bold{\widetilde{\Psi}}(\widetilde{\x})
  =\rho(g^{-1})\widetilde{\x}^{-1}\Bold{\psi}(\x)
  =:\rho(g^{-1})\rho(\widetilde{g})\Bold{\Psi}(\x)
\end{equation}
for some $g\in G$ and fixed $\widetilde{g}\in G$ where
$\Bold{\psi}:\M\rightarrow T_{(p,q)}\M$,
$\Bold{\Psi}:\M\rightarrow (\mathbb{C}^m)^{p+q}$, and $\rho
:G\rightarrow GL((\mathbb{C}^m)^{p+q})$ is a (possibly
non-faithful) representation of $G$. Note that $\widetilde{\x}$ is
both a point in $\widetilde{\M}$ and an admissible map
$\widetilde{\x}:(\mathbb{C}^m)^{p+q})\rightarrow\pi^{-1}(\M)$
where $\pi:T_{(p,q)}\M\rightarrow\M$. Also,
$\rho(\widetilde{g})\Bold{\Psi}$ is, by definition, the tensor
field
$\rho(\widetilde{g})\Bold{\Psi}:=\widetilde{\x}^{-1}\Bold{\psi}$
associated with the equivariant map $\Bold{\widetilde{\Psi}}$. The
$\rho(\widetilde{g})$ factor has been included because there is no
\emph{a priori} relationship between $\Bold{\Psi}(\x)$ and
$\Bold{\widetilde{\Psi}}(\widetilde{\x})$. The elementary and
boundary kernels are required to be equivariant in their arguments
in the above sense.

Consider the paths $x\in\PU$ parametrized by the composition
$P=P_{\Pi}\circ\widetilde{P} $ where $\widetilde{P}:\Za\rightarrow
{\mathit{P}_a^{\widetilde{\mathbb{V}}}\widetilde{\mathbb{U}}}$ and
$P_{\Pi}:{\mathit{P}_a^{\widetilde{\mathbb{V}}}\widetilde{\mathbb{U}}}
\rightarrow {\mathit{P}_a^{\mathbb{V}}\mathbb{U}}$ where
$\widetilde{\mathbb{U}}\subseteq \widetilde{\M}$ and
$\mathbb{U}\subseteq \M$. For a fixed initial point $\x_a$, a path
$x(\cdot ,z)$ can be horizontally lifted by the connection
yielding the family of paths $\{\widetilde{x}_{(p)}(\cdot
,z_{(p)})\}$ indexed by $p$ where $p$ runs over the elements $g_p$
of $G$. The (multi-)index $p$ is discrete or continuous depending
on the nature of $G$. The paths are related by
$\widetilde{x}_{(q)}(\cdot ,z_{(p)})=\widetilde{x}_{(p)}(\cdot
,z_{(p)})\cdot g_{(qp)}$ for some $g_{(qp)}\in G$.

Define the evaluation map $\varepsilon:\Za\rightarrow \mathbb{U}$
and its lift $\widetilde{\varepsilon}:\Za\rightarrow
\widetilde{\mathbb{U}}$. Then
$\widetilde{\varepsilon}=\Pi^{-1}\circ\varepsilon$. Paths with
fixed end-point $\x_{a}$ in the base space are lifted to paths
with fixed end-points in the fiber over $\x_{a}$. It is clear that
the union of the inverse images of the lifted paths
$\widetilde{\varepsilon}^{-1}(\widetilde{x}_{(p)}(\cdot,z_{(p)})$
is mapped to base space paths with fixed end-point $\x_{a}$ by the
evaluation map $\varepsilon$. Hence, a path integral over
${\mathit{P}_a^{\widetilde{\mathbb{V}}}\widetilde{\mathbb{U}}}$
(which is defined in terms of $\Za$) decomposes as the sum (or
integral), over the elements of a discrete (or Lie) group $G$, of
integrals over ${\Za}_p$ where
$\widetilde{x}_{(p)}(\cdot,z_{(p)})\in{\Za}_p$. Under this
decomposition, the elementary kernel on $\widetilde{\M}$ becomes
\begin{eqnarray}
  \widetilde{\K}_\mathbb{U}(\widetilde{\x}_a,\widetilde{\x}_{a'})
  &=&\sum_{p}\!\!\!\!\!\!\!\!\int\int_{\Omega_{p}}\taurange
  \theta(\taubot-\tau_{a'})
  \widetilde{\Bold{\delta}}(\widetilde{x}_{(p)}
  (\tau_{a'},z_{(p)}),\widetilde{\x}_{a'})\nonumber\\
  &&\hspace{.4in}\times
  \exp{\left\{-\mathcal{S}(\widetilde{x}_{(p)}
  (\tau_{a'},z_{(p)}))\right\}}
  \;d\tau_{a'}\,\DO{\Omega_{p}}
\end{eqnarray}
where the summation/integral symbol means $\sum_{p}$ for discrete
or disconnected $G$ and $\int_G dg$ for $G$ the connected
component of a Lie group.

For some fiducial path with end-points
$\widetilde{\x}_{(0)a}\in\Pi^{-1}(\x_a)$ and
$\widetilde{\x}_{(0)a'}\in\Pi^{-1}(\x_{a'})$, then
$\widetilde{x}_{(p)}(\tau_{a'},z_{(p)})
=\widetilde{\x}_{(0)a}\cdot\widetilde{\mathit{\Sigma}}
(\tau_{a'},z_{(0)})\cdot g_p$ and
$\widetilde{\x}_{a'}=\widetilde{\x}_{(0)a'}\cdot g_p\,$. This,
together with the requirement that $\widetilde{\K}$ be equivariant
in its two arguments, implies
\begin{eqnarray}
\widetilde{\Bold{\delta}}(\widetilde{x}(\tau_{a'},z_{(p)}),\widetilde{\x}_{a'})
&=&\widetilde{\Bold{\delta}} (\widetilde{\x}_{(0)a}\cdot
g_p\cdot\widetilde{\mathit{\Sigma}}_p(\tau_{a'},z_{0}),
\widetilde{\x}_{(0)a'}\cdot g_p)\nonumber\\
&=&\rho(g_p^{-1})\widetilde{\Bold{\delta}}
(\widetilde{\x}_{(0)a}\cdot\widetilde{\mathit{\Sigma}}_p(\tau_{a'},z_{0}),
\widetilde{\x}_{(0)a'})
\end{eqnarray}
where
\begin{equation}
  \widetilde{\mathit{\Sigma}}_p(\tau_{a'},z_{0})
  :=g_p^{-1}\cdot \widetilde{\mathit{\Sigma}}(\tau_{a'},z_{0})\cdot
  g_{p}
  =\widetilde{\mathit{\Sigma}}(\tau_{a'},z_{(p)})\;.
\end{equation}
Hence,
\begin{eqnarray}
  \widetilde{\K}_\mathbb{U}(\widetilde{\x}_a,\widetilde{\x}_{a'})
  &=&\sum_{p}\!\!\!\!\!\!\!\!\int\rho(g_p^{-1})\int_{\Omega_p}\taurange
  \theta(\taubot-\tau_{a'})
  \widetilde{\Bold{\delta}}
(\widetilde{\x}_{(0)a}\cdot\widetilde{\mathit{\Sigma}}_p(\tau_{a'},z_{0}),
\widetilde{\x}_{(0)a'})\nonumber\\
  &&\hspace{.4in}\times
  \exp{\left\{-\mathcal{S}(\widetilde{x}_{(p)}(\tau_{a'},z_{(p)}))\right\}}
  \;d\tau_{a'}\,\DO{\Omega_{p}}\;.
\end{eqnarray}
Using (\ref{important relation}), the kernel on $\M$ can be
expressed as
\begin{eqnarray}\label{intermediate result}
  \K_\mathbb{U}(\x_a,\x_{a'})
  &=&\rho(\widetilde{g}^{-1})\widetilde{\K}_\mathbb{U}
  (\widetilde{\x}_a,\widetilde{\x}_{a'})
  \nonumber\\
  &=&\rho(\widetilde{g}^{-1})\sum_{p}\!\!\!\!\!\!\!\!\int
  \rho(g_p^{-1})\int_{\Omega_{p}}\taurange
  \theta(\taubot-\tau_{a'})
  \widetilde{\Bold{\delta}}
(\widetilde{\x}_{(0)a}\cdot\widetilde{\mathit{\Sigma}}_p(\tau_{a'},z_{0}),
\widetilde{\x}_{(0)a'})\nonumber\\
  &&\hspace{.4in}\times
  \exp{\left\{-\mathcal{S}(\widetilde{x}_{(p)}(\tau_{a'},z_{(p)}))\right\}}
  \;d\tau_{a'}\,\DO{\Omega_{p}}\nonumber\\
  &=:&\sum_{p}\!\!\!\!\!\!\!\!\int\rho(g_p^{-1})
  \,\rho(\widetilde{g}^{-1})
  \widetilde{\K}_{\mathbb{U}(p)}
  (\widetilde{\x}_{(0)a},\widetilde{\x}_{(0)a'})\nonumber\\
  &=&\sum_{p}\!\!\!\!\!\!\!\!\int\rho(g_p^{-1})
  \widetilde{\K}_{\mathbb{U}(p)}(\widetilde{\x}_{(0)a},\widetilde{\x}_{(0)a'})
\end{eqnarray}
for some $\widetilde{g}\in G$. In the last line, the
$\rho(\widetilde{g}^{-1})$ was absorbed in a redefinition of the
fiducial end-point (which was arbitrary anyway).

Equation (\ref{intermediate result}) is still not satisfactory
because the term $\rho(g_p^{-1})$ is not explicitly
known.\footnote{However, in quantum mechanical applications where
$\Bold{\Psi}$ represents a wave function, (\ref{important
relation}) implies immediately that $\rho$ must be a unitary
(projective)representation.} Also, there is ambiguity in the
choice of the fiducial path so I could just as well replace $g_p$
by, say, $\overline{g}_p=g_p\cdot g_0$ for some $g_0\in G$.
Proceed by taking the absolute value of (\ref{intermediate
result}) putting $g_p\rightarrow \overline{g}_p$\,;
\begin{eqnarray}
  \left|\overline{\K}_\mathbb{U}(\x_a,\x_{a'})\right|&=&
 \left|\sum_{p}\!\!\!\!\!\!\!\!\int\rho({\overline{g}_p}^{-1})
  \widetilde{\K}_{\mathbb{U}(p)}
  (\widetilde{\x}_{(0)a},\widetilde{\x}_{(0)a'})\right|\;.
\end{eqnarray}
Since the modulus of the kernel should not depend on the choice of
fiducial path, then $|\overline{\K}|=|\K|$. A necessary and
sufficient condition for this to hold is that $\rho$ must furnish
a unitary representation of $G$ (see e.g. \cite{LA/D-M}).

Because the fiducial path is arbitrary, it is convenient to choose
$\widetilde{\x}_{(0)}=(\x,e)$ where $e$ is the identity element in
$G$. With this choice, the elementary kernel on $\mathbb{U}$ is
given by
\begin{equation}
  \K_\mathbb{U}^r(\x_a,\x_{a'})=
  \sum_{p}\!\!\!\!\!\!\!\!\int\rho^r(g_p^{-1})
  \widetilde{\K}_{\mathbb{U}(p)}(\x_a,\x_{a'})
\end{equation}
where the $\rho^r$ furnish unitary representations of the
structure group $G$ and I am justified in writing
\begin{eqnarray}
  \widetilde{\K}_\mathbb{U}(\x_a,\x_{a'})
  &=&\int_{\Omega_{p}}\taurange
  \theta(\taubot-\tau_{a'})
  \widetilde{\Bold{\delta}}
(\x_{a}\cdot\widetilde{\mathit{\Sigma}}_p(\tau_{a'},z_{0}),
\x_{a'})\nonumber\\
  &&\hspace{.4in}\times
  \exp{\left\{-\mathcal{S}(x_{(p)}(\tau_{a'},z_{(p)}))\right\}}
  \;d\tau_{a'}\,\DO{\Omega_{p}}\;.
\end{eqnarray}
where reference to the point $e$ has been omitted since it is
inconsequential.


\end{document}